\documentclass[11pt]{elsarticle}

\makeatletter
\def\ps@pprintTitle{%
 \let\@oddhead\@empty
 \let\@evenhead\@empty
 \def\@oddfoot{}%
 \let\@evenfoot\@oddfoot}
\makeatother

\usepackage{caption} 
\usepackage[hyphens]{url}

\usepackage[margin=1in]{geometry}
\usepackage{setspace}
\usepackage{amsmath,amssymb,amsthm,bm}
\usepackage{graphicx}
\usepackage{multicol}
\usepackage{flafter}
\usepackage{hyperref,xcolor}
\definecolor{ForestGreen}{RGB}{34,139,34}
\usepackage{float}
\usepackage{algorithm2e}
\usepackage{framed}
\hypersetup{colorlinks=true,urlcolor=blue,citecolor=purple}
\usepackage{cleveref}
\usepackage{longtable}
\usepackage{float}
\usepackage[export]{adjustbox}
\usepackage{placeins}
\usepackage{blkarray}
\usepackage{booktabs, bigstrut, multicol}

\biboptions{authoryear}

\numberwithin{equation}{section}

\def\boldeps{\bm{\varepsilon}}

\def\R{\mathbb{R}}

\def\N{\mathbb{N}}
\def\No{\mathcal{N}}

\def\L{\mathcal{L}}

\def\M{\mathcal{M}}

\newcommand{\prob}{\mathbb{P}}

\newcommand{\abs}[1]{\left\lvert #1 \right\rvert}

\newcommand{\cov}{\mbox{Cov}}

\newtheoremstyle{general}
{3mm} 
{3mm} 
{} 
{} 
{\bfseries} 
{.} 
{.5em} 
{} 

\theoremstyle{general}

\begin{document}

\begin{frontmatter}

\title{Real-time forecasting within soccer matches through a Bayesian lens}
\author[IIMB]{Chinmay Divekar}
\author[IIMB]{Soudeep Deb}
\author[UE]{Rishideep Roy}
\affiliation[IIMB]{organization={Indian Institute of Management Bangalore},
	addressline={Bannerghatta Main Road}, 
	city={Bangalore},
	postcode={560076}, 
	state={KA},
	country={India}}
\affiliation[UE]{organization={University of Essex},
	addressline={School of Mathematics, Statistics and Actuarial Science, Wivenhoe Park}, 
	city={Colchester},
	postcode={CO4 3SQ}, 
	state={Essex},
	country={UK}}
\begin{abstract}
This paper employs a Bayesian methodology to predict the results of soccer matches in real-time. Using sequential data of various events throughout the match, we utilize a multinomial probit regression in a novel framework to estimate the time-varying impact of covariates and to forecast the outcome. English Premier League data from eight seasons are used to evaluate the efficacy of our method. Different evaluation metrics establish that the proposed model outperforms potential competitors inspired by existing statistical or machine learning algorithms. Additionally, we apply robustness checks to demonstrate the model's accuracy across various scenarios. 
\end{abstract}


\end{frontmatter}

\section{Introduction}\label{sec:introduction}

Sports and games are recreations that have attracted mankind since time immemorial. Soccer is a significant one among them. Early forms of sports involving feet and balls have been noted in many parts of the world. The game Tsu'chu or Cuju (in China), which literally stands for `kicking the ball', is one of the earliest versions of the game from East Asia and was in practice a few millennia ago. There was a similar sport in Japan called Kemari, which still survives. Similarly, there have been examples of sports using balls and feet from ancient Greek and Roman civilizations, among the Native Americans, the indigenous people of Oceania, etc. There have been similar other sports recorded in Europe, particularly in the modern-day United Kingdom during the Middle Ages. With time, some of the old forms have modified themselves, some have lost out in the race, and others have taken their place. Their popularity, in terms of participation and attracting the audience, keeps the sheen of soccer. As \cite{walvin2014people} notes in his book on the history of soccer, it is literally the ``people's game''.  Spectators are especially engaged in rooting for their favourite heroes and teams and in predicting the outcome of the games. In fact, forecasting and decision-making have become intrinsic parts of sports. For avid fans, betting on their favourite teams provides just as good an experience as watching the game. The global sports betting market is estimated to increase at an approximate compound annual growth rate of 10.3\% till 2032, with soccer (or association football) bringing the most attention, as stated in \cite{SportsCAGR2022}. Arguably the most viewed sport in the world, the sports market connected to soccer is also huge. 

With the increasing viewership and fandom, the concept of within-game result forecasting is turning out to be a very important aspect of soccer. The market associated with this prediction being ever so on the rise, there is a greater requirement on the precision of these predictions, which include not just the result but also the final scoreline. There are numerous techniques to forecast the outcome of a soccer match based on aggregated data at the beginning of a match, but they lack the flexibility to update the predictions based on the sequence of particular events during the match. The fact that different types of events may occur on the soccer field with every passing minute within a match naturally motivates a Bayesian model, where one can estimate the effects of specific types of events as functions of time by taking into account the minute-by-minute data of soccer matches and subsequently use that to predict the final outcome of the game. To the best of our knowledge, there is no published work in this direction. Although many websites, such as \url{https://theanalyst.com/eu/} (for football), \url{https://fivethirtyeight.com/} (for basketball), \url{https://www.dimers.com/} and \url{https://patents.google.com/patent/US20070072679A1/en} (usually for all sports), provide live-win-probability for different matches, they lack the clarity and documentation behind their methods. In the current work, we aim to bridge that gap by developing a predictive Bayesian model that can be used for within-match forecasting in soccer. Our methodology works with the final outcome as an ordinal response variable and models the latent variable by suitably incorporating different covariates and events relevant to the game. On the one hand, the proposed approach allows us to identify the time-varying impacts of various events in soccer, while on the other, it is found to record superior predictive performance than a few other possible approaches, which are developed as extensions of existing statistical and machine learning methods. 

Before delving into the proposed methodology, we find it pivotal to present a succinct account of existing literature that focuses on forecasting problems related to soccer using formal statistical methods, as well as the ones that develop in-game prediction methods for other types of sports.

\subsection{Brief literature review}

Accurate forecasting of results in sports has been an interest of the world for a long time. Quite naturally, due to its popularity and financial impacts, prediction problems in soccer constitute an extensively researched area. Early papers like \cite{maher1982modelling} modelled the distribution of goals using bivariate Poisson random variables. \cite{rue2000prediction} extended the bivariate modelling approach to a Bayesian setup while considering the strength of a team as a time-dependent factor. \cite{crowder2002dynamic} proposed a more computationally efficient method to update the strength parameters. \cite{goddard2005regression} compared the forecasting performance of a goal-based model (bivariate Poisson regression model) against a result-based model (ordered probit model). \cite{mchale2007modelling} extended the idea of dependence in a bivariate framework and utilized copulas to model the results of soccer matches. On the other hand, \cite{hvattum2010using} worked with a rating-based system to predict the result of soccer matches, demonstrating that ELO rating-based measures can adequately incorporate past results. Another advanced technique of sparse bivariate Poisson model, along with the concepts of boosting to select appropriate covariates, was used by \cite{groll2018dependency}. 

Another branch of literature studies the relationship between the match results and bookmaker odds. \cite{forrest2000forecasting} analyzed English soccer match odds provided by newspaper agencies to determine their efficiency. The authors concluded that these odds do not appropriately utilize team strength and such publicly available data to enhance their forecasting prowess. The articles by \cite{vstrumbelj2010online}, \cite{vstrumbelj2014determining} explored the quality of bookmaker odds by viewing them as probabilistic evaluations of the match results. Different aspects of the game that seemed to have an effect on the outcome were also studied. \cite{clarke1995home} explored the effect of home advantage on the game result, while \cite{ley2019ranking} compared the existing team strength-based modelling approaches. \cite{koopman2019forecasting} proposed a dynamic modelling framework for the scenario when the outcome variable is goals scored, win/draw/loss and difference in goals scored. The time-dependent covariates in this method are updated as an autoregressive process. Different machine learning techniques have also been used in this regard. \cite{liti2017predicting} provide an excellent comparative discussion on the accuracy of various such algorithms, while \cite{mendes2020comparing} assess the efficacy of neural network ensemble methods in forecasting match outcomes in soccer. 

Moving on to the literature of in-game forecasting, we note that since the 2000s, with the influx of data, this topic has picked pace in the domain of sports analytics. In cricket, \cite{easton2006examination} demonstrated the impact of in-game ball-by-ball events on the change in the odds of winning. In tennis, \cite{klaassen2003forecasting} modelled the probability of the player serving to win the match and designed a way to update this probability after every point. This model was further explored by \cite{easton2010forecasting} to demonstrate tennis betting markets' in-game efficiency. \cite{stern2005brownian} forecast the winning probability in a game of basketball by modelling the difference in points scored as a Brownian process. An extension of this by incorporating the betting odds and using a Gamma process was proposed by \cite{song2020making}. Intriguingly, even though soccer is arguably the most popular sport across the globe, we could not find any research article to tackle the problem of within-game forecasting in soccer. Our focus in this article is to address this issue through the help of a Bayesian framework.

\subsection{Our contribution}

In sports like basketball, baseball or cricket, the main events deciding the result (points, runs, wickets etc.) typically have a high rate of occurrence; hence a much gradual change is expected to be observed in the outcome probabilities over time. In contrast, the frequency of the corresponding variable (that is, a goal) in professional soccer is extremely low. Consequently, the fate of a soccer match is immensely volatile, which is likely to hinder the accuracy of an in-game forecaster. This article attempts to solve this problem and fill the gap in the extant literature, by proposing a Bayesian model for within-game forecasting in soccer. 

The proposed methodology treats the outcome of a match with respect to the home team as an ordinal multinomial random variable (win/draw/loss). Then, utilizing appropriate time-invariant and time-dependent covariates which record the events happening in real-time in a soccer match, our model forecasts the result as the match progresses. We use a complete Bayesian framework for predicting the final outcome of the match. A latent variable with two cutoffs is used as a proxy for modelling the final outcome. We assume that this latent variable depends on different covariates pertaining to the match, as well as random errors, and this dependence is linear. The latent variable is continuously updated using Bayesian techniques, which in turn predicts the final outcome as well. As we shall discuss in detail below, it is quite a broad framework without making strong assumptions about the model. We use real-life data from the top division of English club football to demonstrate the predictive accuracy of the proposed algorithm. The model is also able to quantify the effects of specific types of events as a function of time during the progression of the match. Furthermore, different robustness checks are done to establish that the method works well across various scenarios.

The outline of the rest of the article is as follows. The methodology used for analysis is described in \Cref{sec:methodology}. We begin with our model specification in detail, followed by the Bayesian techniques used in our work. We also describe a way of evaluating our method, which includes the estimations as well as the predictive distributions. For completeness of the study, we extend a few other existing methods for conventional forecasting in soccer to within-game predictive techniques and compare their performances against our proposed model. \Cref{sec:data} describes the dataset we use for our work, along with some exploratory analysis and descriptive statistics. Our results are demonstrated in \Cref{sec:results} where we present a general analysis, followed by a robustness study. Next, we look at two specific case studies to understand the results better. The article ends with the summary and some important concluding remarks in \Cref{sec:conclusion}.

\section{Methodology}\label{sec:methodology}

\subsection{Model specification}\label{subsec:model}

Throughout this article, we shall use $\N$, $\mathbb{Z}$ and $\R$ to denote the sets of natural numbers, integers and real numbers, respectively. We use the shorthand notation $[K]$ for the set of all natural numbers from $1$ to $K$, i.e.\ $\{ k:1 \leqslant k \leqslant K; k \in \N \}$.  Any matrix is represented in bold capital letters, and any vector in lower-case bold letters; for example, $\bm{a}$ would indicate a vector and $\bm{A}$ would represent a matrix. We shall use $\bm I_d$ to denote an identity matrix of dimension $d\times d$. Also, we shall use $\bm{1}(\cdot)$ to denote an indicator function. A $d$-dimensional normal distribution is indicated by $\mathcal{N}_d(\cdot,\cdot)$ whereas a truncated normal distribution is denoted by the shorthand $\mathcal{T}\mathcal{N}(\mu, \sigma^2, a, b)$, where $\mu$ and $\sigma$ are the mean and standard deviation of the original normal distribution, while the truncation limits are $a$ and $b$, with $a<b$.

Let us formally define the framework now. We model the outcome of a game from the perspective of the team playing at home. The dependent variable is therefore denoted by an ordered multinomial random variable with three categories -- loss, draw or win -- correspondingly denoted by $r\in R = \{-1, 0, 1\}$, respectively. The focus of the model is on forecasting the outcome in real-time, that is, after every minute of the match. Therefore, we define a time index set $\Gamma = \{t:1 \leqslant t \leqslant 90 ;~ t \in \N \}$. At the end of every minute $t \in \Gamma$, we record a set of covariates from minutes $1$ through $t$. Hereafter, we refer to these as time-varying covariates.

To define the main model, let $Y_i$ denote the outcome for the home team in the $i^{th}$ match, for $i\in [n]$. The vector of the outcomes $(Y_1,\hdots,Y_n)$ will be denoted as $\bm{y}$. It is treated along the lines of \cite{greene2003econometric}, for an ordered probit model. We define a latent variable $\Pi_i$ and cut-offs $\delta_1, \delta_2$ such that,
\begin{equation}\label{eq:yspec}
Y_i=
    \begin{cases}
    -1, &\mbox{if } \Pi_i < \delta_1 \\
    ~~0, &\mbox{if } \delta _1\leqslant \Pi_i \leqslant \delta_2 \\
    ~~1, &\mbox{if } \Pi_i > \delta_2. 
    \end{cases} 
\end{equation}

Here, $\Pi_i$ is the value of the latent variable in the $i^{th}$ match, which can be modelled as a function of the covariates. It is important to observe that we use 90 different models for time points $t \in \Gamma$. For the model till time $t$, for the $i^{th}$ match, the response is an approximation of the latent variable till time $t$, $\Pi_i^{(t)}$. In this model, we use information of the events on the field till the time point $t \in \Gamma$. Note that \cite{goddard2005regression} modelled the ordinal outcome using probit regression with linear and additive functional forms of the covariates. \cite{angelini2017parx}, \cite{koopman2019forecasting} also follow a similar specification in their papers. Motivated by these studies, we use a linear function with additive error for modelling the outcome of a soccer match in a similar setup. Mathematically, for the $i^{th}$ match, till the time point $t$, $\Pi_i^{(t)}$ is modeled as
\begin{equation}\label{eq:pispec}
    \Pi_i^{(t)} = \textbf{z}_i^{\top} \bm{\gamma} + \sum_{k \in [K]} {\textbf{x}_{ik}^{(t)}}^\top \bm{\beta}_{k}^{(t)} + \boldeps_i^{(t)}, 
\end{equation}
where $\textbf{z}_i$ is a $p\times 1$ vector of time-invariant covariates such as the strengths of the starting elevens for both the teams and $\bm{\gamma}$ is the corresponding vector of coefficients. On the other hand, $\bm{\beta}_k^{(t)}$ is a $2t \times 1$ vector of coefficients capturing the time-varying effect of an event of type $k$ on the outcome of a match, whereas
\begin{equation*}
    \textbf{x}_{ik}^{(t)} = \begin{bmatrix} x_{ik,1}^H & x_{ik,2}^H & \cdots & x_{ik,t}^H & x_{ik,1}^A & x_{ik,2}^A & \cdots & x_{ik,t}^A  \end{bmatrix}^\top, \; \text{for } i \in [n],
\end{equation*}
is the associated vector indicating the number of occurrences for the $k^{th}$ type of event until time $t$ for both teams. Here, $x_{ik,t}^H$ and $x_{ik,t}^A$ denote the corresponding covariates for the home ($H$) and away ($A$) teams, respectively. We assume that there are a total of $K$ types of such events. For instance, one may consider goals, fouls, cards, shots-on-goal, corners etc. The variables used in this article are further elaborated in \Cref{sec:data}. It should be reiterated that the proposed model incorporates all time-varying events in a linear additive fashion.

Finally, we assume that all $\boldeps_i^{(t)}$ are independent and identically distributed (iid) zero-mean Gaussian random variables,  and we define $\boldeps^{(t)} = (\boldeps_1^{(t)}, \ldots, \boldeps_i^{(t)}, \ldots, \boldeps_n^{(t)})^{\top}$. Such an assumption for the error distribution is a popular choice in the current context, see \cite{koning2000balance} for example. \cite{koopman2019forecasting} also explored various methods to predict match results, and used Gaussian marginals since it leads to a parsimonious model with straightforward estimation procedures. Another motivation for such an assumption is that in a Bayesian setting which we use for implementation, a Normal prior for $\boldeps^{(t)}$ induces conjugacy in the posterior distributions and is hence simpler to implement in  Markov chain Monte Carlo (MCMC) methods. With this assumption, we can write 
\begin{equation}\label{eq:epsi}
    \boldeps^{(t)} \sim \No_n(0, \sigma_y^2 \bm{I}_n),
\end{equation}
where $\sigma_y^2$ is the error variance. Due to the structure of the latent variable in the model, we cannot estimate the cutoffs $\bm{\delta}$ and the error variance $\sigma^2_y$ simultaneously (see \cite{higgs2010clipped} for relevant discussions). Thus, we fix $\sigma_y^2$ at around 200 in the estimation algorithm below.

\subsection{Bayesian Estimation}

We use a complete Bayesian framework to estimate the model parameters and provide a probabilistic forecast for the response variable. In that regard, suitable prior specifications are necessary and we elaborate on them below. Also, we shall be using $f(\cdot)$ to denote conditional and marginal densities in the likelihood and prior, and $\pi(\cdot)$ to denote the posterior likelihoods. 

A Gaussian prior for the covariates in a probit model is a well researched topic, cf.\ \cite{mcculloch2000bayesian}. Such a prior has been found to work well in predicting soccer match outcomes too \citep{rue2000prediction}. Accordingly, we specify suitable Gaussian distributions as conjugate priors for $\bm\gamma$ and $\bm\beta_k^{(t)}$ which result in straightforward conditional posterior distributions. In particular, for the two parameter vectors in the mean structure of the model, we consider
\begin{equation}
    \bm{\gamma} \sim \No_p(\bm{0},\bm{\Sigma}_\gamma),
\end{equation}
where $\bm{\Sigma}_\gamma$ is a diagonal matrix, and 
\begin{equation}\label{eq:beta-distribution}
    \bm{\beta}_k^{(t)}= \begin{bmatrix} \beta_{k,1}^H & \beta_{k,2}^H & \cdots & \beta_{k,t}^H &  \beta_{k,1}^A & \beta_{k,2}^A & \cdots & \beta_{k,t}^A \end{bmatrix}^T \sim \No_{2t}(\bm{0}, \bm{\Sigma}_k^{(t)}),
\end{equation}
defined for each $k \in [K]$. Recall that $\bm\gamma$ is the vector of coefficients representing the effect of time-invariant covariates, whereas $\beta_{k,j}^H$ and $\beta_{k,j}^A$ (for $1 \leqslant j \leqslant t$) are the coefficients capturing the time-varying effects for the home ($H$) and away ($A$) teams respectively. In the above formulation, $\bm{\Sigma}_k^{(t)}$ is a $2t \times 2t$ covariance matrix entailing the dependence between the effects of the same type of event with respect to various time points. We now specify the structure of $\bm{\Sigma}_k^{(t)}$, assumed to be common for all events $k \in [K]$. First, we assume that there is no correlation between home ($H$) and away ($A$) team events, that is,
\begin{equation}\label{eq:beta-home-away-indep}
    \cov \left(\beta^H_{k,t_1},\beta^A_{k,t_2}\right) = 0 \; \text{for all} \; t_1,t_2 \in \Gamma.
\end{equation}

Next, to define the covariance structure common to both home and away coefficients, we assume that the dependence between occurrences of event $k$ at distinct time points $t_1$ and $t_2$ is governed by the structure
\begin{equation}
    \cov \left(\beta_{k,t_1}^H,\beta_{k, t_2}^H\right) = \cov \left(\beta_{k,t_1}^A,\beta_{k, t_2}^A\right) = g\left(\abs{t_1 - t_2}\right),
\end{equation}
where $g:\mathbb{R}\to \mathbb{R}$ is a decreasing function. Throughout this paper, we assume an exponential function for $g$, i.e.\ $g(x) \propto e^{-x}$.  We also assume that distinct events $k, k' \in [K]$ are not correlated amongst themselves. 

Finally, we specify the prior distributions for the cutoffs $\delta_1,\delta_2$. Previous works such as \cite{albert1993bayesian} and \cite{liddell2018analyzing} demonstrate that Gaussian priors work well for the cut-offs in a probit model. Taking motivation from these studies, we assume
\begin{equation}\label{deltas}
    \delta_j \sim  \No(0,\tau^2), \; j \in \{ 1,2 \},
\end{equation}
where $\tau$ is a large number to ensure high variability of the prior. For the purpose of this paper, we shall always use $\tau=200$.

In order to explain the main steps of the Bayesian estimation procedure, one can note that the set of time-varying and time-invariant covariates can be combined into a single matrix denoted by $\bm{\M}^{(t)}$, and the proposed model in equation (\ref{eq:pispec}) can be rewritten as
\begin{equation}
     \Pi^{(t)} = \left(\bm{\M}^{(t)}\right)^{\top} \bm{\nu}^{(t)} + \bm{\varepsilon^{(t)}}, 
\end{equation}
where $\Pi^{(t)}$ is the vector of outcomes of dimension $n \times 1$, and $\boldeps^{(t)}$ is defined the same as in \cref{eq:epsi}. Similarly, $\bm{\M^{(t)}}$ is a matrix of dimension $(p+2Kt) \times n$ comprising of all covariates, while $\bm{\nu}^{(t)}$ is a $(p+2Kt) \times 1$ vector of parameters to be estimated. Since both $\bm{\gamma}$ and $\bm{\beta}_k^{(t)}$ are assumed to have Gaussian priors, we can further write 
\begin{equation}
    \bm{\nu}^{(t)} \sim \No_{p+2Kt}\left(\bm{0}, \bm{\Sigma}_0^{(t)}\right), \; \text{where} \;
    \bm{\Sigma}_0^{(t)} = \left[\begin{array}{c c c c}
    \bm{\Sigma}_\gamma & \bm{0} & \hdots & \bm{0} \\
    \bm{0} & \bm{\Sigma}_1^{(t)} & \hdots & \bm{0} \\
    \vdots & \vdots & \ddots & \vdots  \\
    \bm{0} & \bm{0} & \hdots & \bm{\Sigma}_K^{(t)} \\
    \end{array}\right]_{(p+2Kt) \times (p+2Kt)}.
\end{equation}

For implementation purposes, we use the concepts of Gibbs sampling in this work. Recall that the Gibbs sampling scheme is a modification of the conventional Metropolis-Hastings algorithm to obtain a sample from multivariate distributions without a closed form. Interested readers are referred to the works of \cite{geman1984stochastic} and \cite{gelfand2000gibbs} for further reading on this algorithm, which relies on the principle that iterative samples from the conditional posterior distributions will lead to a sample representative of their joint distribution. We shall now be dropping the time index $t$ for the purposes of simplicity. We follow the Gibbs sampling procedure to obtain the joint posterior distribution of $\bm{\nu}$ and $\bm{\delta}$. 

To start with, the likelihood can be written as

\begin{equation}
    \mathcal{L}(\bm{\nu},\bm{\delta}\mid \bm{\M},\Pi, \textbf{y}) = \prod_{i=1}^n \prod_{j\in \{-1, 0, 1\}} \prob(Y_i = j)^{\bm{1}(Y_i = j)}.
\end{equation}

This can equivalently be expressed as 
\begin{equation}
    \mathcal{L}(\bm{\nu},\bm{\delta}\mid \bm{\M},\Pi, \textbf{y}) = \prod_{i=1}^n  \prob(\Pi_i < \delta_1)^{\bm{1}(\Pi_i < \delta_1)}\times \prob(\delta_1 \leqslant \Pi_i \leqslant \delta_2)^{\bm{1}(\delta_1 \leqslant \Pi_i \leqslant \delta_2)} \times \prob(\Pi_i > \delta_2)^{\bm{1}(\Pi_i > \delta_2)}.
\end{equation}

Recall that $\Pi_i$ is the latent variable in the model. Moreover, for the proposed structure, we can equivalently denote $Y_i$ by an expression with $\Pi_i$ and $\bm{\delta}$. Since we employ a Gibbs sampling procedure with the latent variable, we restrict ourselves to the use of $\Pi$ and $\bm\delta$ for the estimation procedure instead of $Y_i$. For computational purposes, we treat ${\Pi}_i$ as the parameter vector, and we compute the conditional distribution of it for the Gibbs sampling steps. It is straightforward to note that the conditional posterior of ${\Pi}_i$ is given by: 
\begin{equation}
    \Pi_i \mid \bm{\M}, \bm{\nu}, \bm{\delta} \sim \mathcal{T}\mathcal{N} (  \bm{\M}_{i}'\bm{\nu}, \sigma_y^2, \delta_{j-1}, \delta_j ), 
\end{equation}
where $\delta_{j-1}$ and $\delta_j$ are the truncation limit for an observation $i$ dependent on the outcome $Y_i$. We define $\delta_0 = -\infty$, $\delta_3 = \infty$ for the sake of completeness.
Next, the choice of a Gaussian prior assumed for $\bm{\nu}$ is essential and leads to a simple closed form expression for its conditional posterior, which can be written as
\begin{equation}
    \pi(\bm{\nu}\mid \bm{\M},\Pi,\textbf{y}, \bm{\delta}) \propto \prod_{i=1}^n f(\Pi_i, \bm{\delta} \mid \bm{\nu},\bm{\M}) f(\bm{\nu}).
\end{equation}

From our assumption of a Gaussian prior distribution, we get
\begin{equation}
    \pi(\bm{\nu}\mid\bm{\M},\Pi,\textbf{y}, \bm{\delta}) \propto \exp \left[ -\frac{1}{2\sigma^2_y} \sum_{i=1}^n (\Pi_i - \M_i^{\top}\bm{\nu})^2 \right] \exp \left[ -\frac{1}{2} \bm{\nu}' \bm{\Sigma}^{-1}_0 \bm{\nu} \right].
\end{equation}

After some algebraic simplification and comparing it to a Gaussian distribution, we observe that the conditional posterior distribution of $\bm{\nu}$ can be expressed as,
\begin{equation}
  \bm{\nu}\mid\bm{\M},\Pi,\textbf{y}, \bm{\delta} \sim \No_{p+2Kt} \left( \bm{\mu}_\nu, \bm{\Tilde{\Sigma}} \right), \label{postnu}
\end{equation}
where $\bm{\Tilde{\Sigma}} = (\sigma^{-2}_y \bm{\M}'\bm{\M} + \bm{\Sigma}^{-1}_0)^{-1}$ and $\bm{\mu}_\nu = \sigma^{-2}_y \bm{\Tilde{\Sigma}}(\bm{\M}'\Pi)$. 
 
Now, to estimate $\bm{\delta}$, we refer to the work of \cite{dechi2019bayesian}. The author establishes a correspondence between the multinomial categories through a transformation of the Dirichlet distribution. The Dirichlet distribution is a multivariate generalization of the Beta distribution with the support being a set of $l$-dimensional vectors with non-negative entries $\L$ such that $||\L||_1$ is one. These can be considered as probabilities of a multinomial outcome with $l$ categories. It is a natural conjugate for a multinomial outcome. For this paper, we assume a Dirichlet ($\alpha_1,\alpha_2,\alpha_3$) distribution with each $\alpha_i = 1$. We choose this specification since it makes for a non-informative prior and due to the favourable convergence properties of its conditional marginals. The relationship between the Dirichlet parameters and the cutoffs $\delta_i$ is defined as, 
\begin{equation}
    \prob(\delta_i \leqslant \delta < \delta_{i+1}) = p_{i+1},
\end{equation}
where $p_{i}$ is the support of the distribution such that $\sum_{i=1}^3 p_i = 1$, and $p_i \in [0,1] \hspace{0.2cm} \forall$ $i \in \{ 1,2,3 \}$. Recall that we have defined $\delta_0 = -\infty$, $\delta_3 = \infty$ before. Then, the joint distribution of $\bm{\delta}$ can be written as,
\begin{equation}
    f(\bm{\delta}) = F(\delta_1)^{\alpha_1 - 1} \left[F(\delta_2) - F(\delta_1)\right]^{\alpha_2 - 1} [1 - F(\delta_2)]^{\alpha_3 - 1}\prod_{j=1}^{2}f(\delta_j),
\end{equation}
where $F(.)$ is any cumulative distribution function with $\R$ being the domain of the random variable. Simple algebraic manipulation leads to the joint conditional posterior likelihood for $\bm{\delta}$ being expressed as, 
\begin{equation}
    \pi(\bm{\delta}\mid\bm{\M},\Pi,\textbf{y},\bm{\nu}) \propto f(\boldsymbol{\delta}) \prod_{i=1}^{n} \prod_{j=1}^{3} \bm{1}(\delta_{j-1} < \Pi_i < \delta_j).
\end{equation}

We now convert the joint likelihood into marginal by conditioning on the other $\delta_j$'s. Because of the Gaussian prior for $\bm{\delta}$ in \cref{deltas}, the conditional posterior distribution for each $\delta_j$ is simply
\begin{equation}
    \pi\left(\delta_j \mid \bm{\M},\Pi,\textbf{y},\bm{\nu},\delta_{-j}\right) \propto \left[\Phi(\delta_j) - \Phi(\delta_{j-1})\right]^{\alpha_j - 1} \left[\Phi(\delta_{j+1} - \Phi(\delta_j))\right]^{\alpha_{j+1}-1} \phi(\delta_j) \bm{1}(c_{j,1} <\delta_j < c_{j,2}) \label{condpostdelta},
\end{equation}
where $c_{j,1} = \max\{\Pi_i:Y_i=j\}$, $c_{j,2} = \min\{\Pi_i:Y_i = j+1\}$ for $j = 1,2$, and $\Phi(.)$ is the cumulative distribution function (CDF) of a univariate Gaussian distribution. For our model with three categories, we define $\Phi(\delta_0) = 0$ and $\Phi(\delta_3) = 1$ for completeness. We now identify the conditional posterior CDF for $\delta_j|\delta_{-j}$. In that regard, we define $\omega = \left(\Phi(\delta_j) - \Phi(\delta_{j-1})\right)/\left(\Phi(\delta_{j+1}) - \Phi(\delta_{j-1})\right).$ Letting $\Phi(\delta_{j-1}) = a$ and $\Phi(\delta_{j+1}) = b$, we can write $\delta_j = \Phi^{-1}((b-a)\omega +a)$. Utilizing the distribution of $\delta_j$, with regard to the distribution of $\omega$, it can be defined as  $f(\omega) = \pi_{\delta}(\delta_j) \frac{\partial \delta_j}{\partial \omega}$,
where $\pi_\delta(.)$ is the unconditional posterior distribution of $\delta_j$.  By substituting the form of $\delta_j$ in \cref{condpostdelta}, we can obtain the expression 
\begin{equation}
    \pi_{\delta}\left(\Phi^{-1}((b-a)\omega + a)\right) \propto \left(\Phi(\delta_j) - a\right)^{\alpha_j-1} \left(b - \Phi(\delta_j)\right)^{\alpha_{j+1}-1}\phi(\delta_j).
\end{equation}
A simple algebraic simplification of the above equation and ignoring terms without $\omega$ leads us to,
\begin{equation}
        \pi_\delta(\delta_j) \propto \omega^{\alpha_{j}-1}(1-\omega)^{\alpha_{j+1}-1}\phi(\delta_j). \label{postdelta}
\end{equation}
The derivative of $\omega$ with respect to $\delta_j$ is simply given by $\frac{\partial\omega}{\partial\delta_j}= \frac{b-a}{\phi(\delta_j)}$. 
Now that we have obtained simplified expressions for the terms required to identify $f(\omega)$, one can argue that
\begin{equation}
    f(\omega) \propto \omega^{\alpha_{j}-1}(1-\omega)^{\alpha_{j+1}-1}\phi(\delta_j) \times \frac{b-a}{\phi(\delta_j)}. 
\end{equation}
Simplifying the above expression and comparing it to a Beta distribution, we get
\begin{equation}
    \omega \sim \mathrm{Beta}(\alpha_j,\alpha_{j+1}). \label{omegaa}
\end{equation}
In this manner, we can now incorporate the parameters of the Dirichlet distribution, namely $\alpha_j$, in the distribution of $\delta_j$ through a closed-form expression for $\omega$. Hence, the conditional posterior distribution of $\delta_j$ can be expressed as
\begin{equation}
\Phi(\delta_j \mid \bm{\M},\Pi,\textbf{y},\bm{\nu}, \delta_{-j}) \sim [\Phi(\delta_{j+1}) - \Phi(\delta_{j-1})] \mathrm{Beta}(\alpha_j,\alpha_{j+1}) + \Phi(\delta_{j-1}).
\end{equation}

Note that $\Phi(\delta_j)$ will be truncated from below and above by $\Phi(c_{j,1})$ and $\Phi(c_{j,2})$ respectively. Let us use the shorthand notation $\Phi_{\delta_j}$ to denote the conditional distribution of $\delta_j$ given the other cutoffs. Therefore, in the modelling setup above, in the case of three ordered categories we obtain the following set of conditional posterior distributions:
\begin{equation}\label{postdeltafinal}
\begin{split}
    \Phi(\delta_1 \mid \delta_2,\bm{\M},\Pi,\textbf{y},\bm{\nu}) &\sim \Phi(\delta_2) \mathrm{Beta}(\alpha_1,\alpha_2) \hspace{3.5cm} \Phi(c_{1,1}) \leqslant \Phi_{\delta_1} \leqslant \Phi(c_{1,2}), \\
    \Phi(\delta_2 \mid \delta_1,\bm{\M},\Pi,\textbf{y},\bm{\nu}) &\sim [1 - \Phi(\delta_1)] \mathrm{Beta}(\alpha_2,\alpha_3) + \Phi(\delta_1) \hspace{1.25cm} \Phi(c_{2,1}) \leqslant \Phi_{\delta_2} \leqslant \Phi(c_{2,2}).
\end{split}
\end{equation}

Now that we have the closed-form expressions for all the conditional posterior distributions corresponding to the unknown parameters in the model, we can implement the Gibbs sampling algorithm to sample from the joint posterior distribution. Following the principles of this algorithm, we need to sequentially sample from the distributions of $(\Pi \mid \bm{\nu},\bm{\delta})$, $(\bm{\nu} \mid \Pi,\bm{\delta})$ and $(\delta_j \mid \delta_{-j},\Pi,\bm{\nu})$ until convergence. We use the GW statistic to assess the convergence, and it is explained below. After the Markov chains in the Gibbs sampler converge, we can obtain a sample from their joint distribution. To ensure the independence of the observations we incorporate thinning by drawing samples from every $10^{th}$ iteration.  The steps followed in the implementation of the Gibbs sampling procedure are now summarized in Algorithm \ref{alg:one}. 

\RestyleAlgo{ruled}
\begin{algorithm}[h!]
\caption{Gibbs sampler for the posterior distribution in the proposed model}\label{alg:one}
\textbf{Input}: Dataset ($\textbf{y},\bm{\M}$) where, \textbf{y} is the multinomial outcome variable and $\bm{\M}$ is the set of covariates. \\
\textbf{Output}: A sample of size S from the posterior distribution of the tuple $(\Pi_i,\bm{\nu},\bm{\delta})$ \\
\textbf{Initialize}: $\Pi^{(0)}$, $\bm{\nu}^{(0)}$ and $\bm{\delta}^{(0)}$. Let $\mathcal{C}$ be the convergence criteria. \\
Let $m \gets 1$ \\
\While {$\mathcal{C}$ not met} 
{ \vspace{0.2cm}  Sample from $\Pi_i^{(m)} \mid \bm{\nu}^{(m-1)},\bm{\delta}^{(m-1)}  \sim \mathcal{T}\No(\bm{\M}_i^{\top}\bm{\nu}^{(m-1)},\sigma^2_y, \delta_{j-1}, \delta_j) $ \\ \vspace{0.3cm} 
Sample from $\bm{\nu}^{(m)} \mid \Pi^{(m)},\bm{\delta}^{(m-1)} \sim \No(\Tilde{\bm{\nu}}^{(m)},\bm{\Tilde{\Sigma}})$, where $\Tilde{\bm{\nu}}^{(m)}$ is the updated mean of the conditional posterior distribution after observing $\Pi^{(m)}$ \\ \vspace{0.3cm} 
\ForEach {$j \in \{1,2\}$} {\vspace{0.2cm} Sample from  \vspace{0.2cm} \\
$\Phi(\delta_j^{(m)}) \mid \bm{\delta}_{(-j)}^{(m-1)},\Pi^{(m)},\bm{\nu}^{(m)} \sim [\Phi(\delta_{j+1}^{(m-1)}) - \Phi(\delta_{j-1}^{(m)})] \mathrm{Beta}(\alpha_j,\alpha_{j+1}) + \Phi(\delta_{j-1}^{(m)})$  \vspace{0.3cm}\\
with, $\Phi(\delta_j^{(m)}) \in [\Phi(c_{j,1}), \Phi(c_{j,2})]$ \\
\vspace{0.3cm}} Let $m \gets m + 1$}
\normalfont{Discard these first M iterations (until $\mathcal{C}$ is met) as burn-in sample}. \\
Continue the iteration procedure given above until we reach iteration $M+S$. The requisite sample is obtained from iterations $M+1$ to $S$.
\end{algorithm}



In order to assess the convergence of the Gibbs sampler, we use the \cite{geweke1991evaluating} statistic, hereafter abbreviated as GW statistic. In this diagnostic approach, a single chain is generated using the Gibbs sampler, and the spectral analysis is used to assess the convergence of the procedure. To elaborate, let $\{g(\theta^{(1)}), g(\theta^{(2)}), \dots g(\theta^{(n)}) \}$ be the iterations of the Gibbs sampler and $S_g(w)$ be the spectral density for the series. Then, under regularity conditions, we can write,
\begin{equation}\label{eq:Geweke-diag}
    E(g(\theta)) = \Bar{g}_n = \frac{1}{n} \sum_{i=1}^n g^{(\theta^{(i)})}, 
\end{equation}
with the asymptotic variance being $S_g(0) / n$. After $n$ iterations of the Gibbs sampler, the GW statistic is calculated by taking the standardized difference between the means $\Bar{g}_n^A$ and $\Bar{g}_n^B$,  which are computed based on the first $n_A$ and the last $n_B$ iterations respectively. As $n$ increases, this statistic tends to a standard normal distribution \citep{cowles1996markov}, and that property is utilized to find evidence of the convergence.

\subsection{Predictive analysis}
\label{subsec:comparison}

It is important to note that the methodology developed above is primarily used to forecast the match outcome after every time-point. To explain this within-game forecasting technique, consider that the model needs to be trained on the dataset $\mathcal{S}_{tr}$, and let the test dataset be $\mathcal{S}_{te}$, with $\abs{\mathcal{S}_{te}}=m$. We shall compare the evaluation metrics (to be elaborated below) for our model as well as for other potential models (discussed later) using the test data. Suppose, we wish to obtain the in-game predictions for the $i^{th}$ match in $\mathcal{S}_{te}$. We shall compute this as a function of $t\in \Gamma$. We know that data are recorded on various events such as goals, corners, cards etc., which take place every single minute. This renders a minute-by-minute record of events which have occurred till time $t$. Following the same notations as before, for the training data, let the set of covariates after time $t$ be denoted as $\bm{\M}$, and the entire training data as $\mathcal{D} = \{\bm{\M},\textbf{y},\Pi\}$. Our aim is to implement the proposed model on $\mathcal{D}$ to predict the outcome of the $i^{th}$ match in $\mathcal{S}_{te}$, after $t$ minutes have passed in the match. Let us use $\hat{\Pi}_i^{(t)}$ to denote the estimated latent variable in this regard, and define the vector $\hat{\Pi}^{(t)} = [\hat{\Pi}_1^{(t)},\dots,\hat{\Pi}_m^{(t)}]^{\top}$ which furnishes the forecasts for all matches in the test dataset at time $t \in \Gamma$. 

Let us use the general notation $\bm{\M_*} = [\bm{\M_{1*}},\dots,\bm{\M_{(p+2Kt)*}}]^{\top}$ to denote the set of covariates corresponding to the test set. Then, $\hat{\Pi}^{(t)}_i$ can be obtained from the posterior predictive distribution that estimates the probabilistic structure of the outcome variable given a new set of covariates. It incorporates the variability in the parameters by weighting the likelihood of $\hat{\Pi}^{(t)}$ by the posterior distribution of the parameters. For our model, we can write
\begin{equation}\label{eq:posterior-predictive}   
\pi\left(\hat{\Pi}^{(t)} \mid \bm{\M_*},\mathcal{D}\right) = \int_{\bm{\nu}} f\left(\hat{\Pi}^{(t)} \mid \bm{\M_*},\mathcal{D},\bm{\nu}\right) \pi(\bm{\nu} \mid \mathcal{D}) \, d\bm{\nu}.
\end{equation}

Here, the distribution of the forecasts for the test data is conditional on the training data $\mathcal{D}$ and the posterior distribution of $\bm{\nu}$. The first term in the expression is simply a likelihood taking a Gaussian form, due to \cref{eq:epsi}. The second term is the posterior distribution of $\bm{\nu}$ given by \cref{postnu}. The above expression can be simplified since both are normally distributed. Thus, the posterior predictive distribution for $\hat{\Pi}^{(t)}_i$ can be simplified as
\begin{equation}
    \pi\left(\hat{\Pi}^{(t)}_i \mid \bm{\M_{i*}},\mathcal{D}\right) \sim \mathcal{T}\No \left(  \bm{\M}_{i*}^{\top} \bm{\mu}_\nu, (\sigma^{2}_y + \bm{\M}_{i*}^{\top} \bm{\Tilde{\Sigma}} \bm{\M}_{i*}) , \delta_{j-1}, \delta_j\right), \label{postpred}
\end{equation}
where $\bm{\Tilde{\Sigma}}$ is the posterior covariance matrix of $\bm{\nu}$, and $\delta_{j-1}$ and $\delta_j$ are the truncation limits. As stated above, the estimate of the latent variable, denoted by $\hat{\Pi}$, for the training set can be obtained from the posterior sample through the Gibbs sampler outlined in Algorithm \ref{alg:one}. We also estimate $\bm{\Tilde{\Sigma}}$, and the cutoffs $\hat{\bm\delta}$ from the posterior samples. Now, in order to predict the outcome, one can obtain a sample from the above posterior predictive distribution, and use that along with the cutoffs to get $\hat{Y}_i^{(t)}$ as the predicted category based on the data up to time $t$. Moreover, the probabilities of the different categories for the outcome variable, corresponding to the home team, can be calculated as
\begin{equation}\label{eq:predicted-prob}
\begin{split}
    &\prob\left(\text{Win}\right) = 1- \Phi_{\hat{\Pi}_i}(\hat{\delta}_2), \\
    &\prob\left(\text{Draw}\right) = \Phi_{\hat{\Pi}_i}(\hat{\delta}_2) - \Phi_{\Pi_i}(\hat{\delta}_1), \\
    &\prob\left(\text{Loss}\right) = \Phi_{\hat{\Pi}_i}(\hat{\delta}_1),
\end{split}
\end{equation}
where $\Phi_{\hat{\Pi}_i}$ is the Gaussian cumulative distribution function of $\hat{\Pi}_i$. We are going to use $\hat{Y}_i^{(t)}$ and the derived probabilities in \cref{eq:predicted-prob} to evaluate the forecasting accuracy of the proposed methodology. It is important to reiterate that the evaluation metrics will be computed for the vector of $\hat{\Pi}_i^{(t)}$ for every time point $t \in \Gamma$. 

While the above procedure renders a point forecast for the outcome probabilities, it also gives us an opportunity to compute a credible interval for the forecasts. Due to the Bayesian framework and \Cref{alg:one}, we have the joint posterior distribution of $\bm{\nu}$, $\bm{\delta}_1$ and $\bm{\delta}_2$ at hand. Drawing a sample from said distribution will result in a unique point estimate for $\hat{\Pi}_i^{(t)}$ and consequently $\hat{Y}_i^{(t)}$ and the predicted probability for each category by means of \cref{eq:predicted-prob}. Repeating this procedure many times, we can generate a set of point estimates which can be used to obtain a credible interval for the predicted outcomes. For implementation of this approach, in this paper, we shall draw 1000 samples from the joint posterior distribution of $\bm{\nu}$ and $\bm{\delta}$, and obtain 1000 realizations for $\prob(\text{Win})$, $\prob(\text{Draw})$ and $\prob(\text{Loss})$, as mentioned above. Then, we compute a 95\% equal tailed credible interval for these probabilities. We repeat the process for each model to obtain credible intervals for every time-point. A flavour of this procedure can be found in \Cref{subsec:specific-examples}, wherein we obtain the credible intervals for the predicted probabilities throughout the course of the match, for specific games. 

In the main application of this article, for the completeness of the study, we are going to compare our model with a few other potential approaches. We highlight that there is no existing work on the topic of within-game forecasting in soccer, but we rely on different algorithms that have been used for predictive modelling in different capacities and extend them to develop competing approaches in our context. The first one in this regard is a version of the generalized linear model (GLM). We train the GLM in a standard probit regression framework by considering the data on the available covariates up to time $t$ for all matches, and predict the outcome in the test set based on that. Thus, on the same lines as our proposed method, the GLM needs to be trained after every time point $t \in \Gamma$, and the forecasts are updated accordingly. 

We next refer to the work of \cite{baboota2019predictive}, who identified a set of features which highly influence the result of a soccer match, through extensive feature engineering and selection. The authors used support vector machine (SVM) and random forest algorithms for predicting the outcome of the match. Although not used in a temporal within-game setting, we consider these models as competitors to our model due to their flexibility in implementation. The covariates selected for modelling by the authors are similar to ours, which serve as another motivation for choosing them as competitors. Note that SVMs are very flexible supervised-learning models, usually used for classification problems. In the comparison study below, when we employ the SVM as one of the contenders, we employ two types of SVMs with respect to the kernel used for modelling, which is one of the most important hyper-parameters in this algorithm. For the first model, we assume linearly separable classes and we shall denote it as Linear SVM below. In the second case, a Gaussian radial basis function is used to incorporate non-linearity in the model. This model will be abbreviated as R-SVM hereafter. The R library \texttt{caret} is used for tuning both models. 

As the fourth model in the comparative discussions, we modify the standard random forest (RF) algorithm, typically used in various classification problems related to soccer. This tree-based algorithm is widely implemented due to its tendency to incorporate dependencies between the covariates in the model. The performance of the random forest algorithm usually depends on finding the precise hyper-parameter. We employed a grid search technique for tuning the hyperparameters. We use the functions provided in the \texttt{base} library for the same. The grid search is performed for identifying the $\mathcal{L}_1$ and $\mathcal{L}_2$ regularization parameters, subsample ratio of columns when constructing each tree and hyper-parameters to control for extreme class imbalance. We search over the interval [0,1] for the optimal subsample ratio and [0,5] for the remaining hyper-paramters. The function \texttt{expand.grid} is used, which primarily iterates the model over all possible combination and ranges of the parameters as required \citep[see][for a detailed discussion]{chambers2017statistical}. As the set of possible features in this algorithm, akin to the other contending approaches, we use the same combination of covariates, the information being available up to time $t$ in every match. 

In order to compare the performances of different algorithms mentioned above, we first rely on the F1-score \citep[see Ch.\ 7 of][for more discussions on this]{van1979information}. It is a standard evaluation criterion of prediction accuracy for a categorical outcome variable, and is essentially the harmonic mean of sensitivity (recall) and specificity (precision). It ranges from zero to one, the latter depicting better accuracy. Below, we define this measure for our proposed model, and a similar computation will be done for the other competitors as well. To avoid notational jargon, we also remove the superscripts indicating the forecast made at time $t$, and it is understood throughout that the metrics are computed as a function of time during the progression of the match.

Let $\hat{\Pi}_{i}$ be the predicted value of the latent variable for the $i^{th}$ match. Following the definition of the categorization, we can determine $\hat{Y}_{i}$ as $j$ if $\hat{\delta}_{j-1} \leqslant \hat{\Pi}_{i} < \hat{\delta}_j$. Then, for the $j^{th}$ outcome category $\mathcal{C}_j$, the confusion matrix for the test set is reduced to a $2 \times 2$ matrix given by
\begin{equation}
\begin{blockarray}{cccc}
 & & \mathcal C_j & \mathcal C_{\neq j}  \\
\begin{block}{cc[cc]}
& \mathcal C_j & a_{j,11}  & a_{j,12} \bigstrut[t] \\
& \mathcal C_{\neq j} & a_{j,21} & a_{j,22} \bigstrut[b]\\
\end{block}
\end{blockarray},
\end{equation}
where $a_{j,11}$ indicates the total number of matches in the test set where correct classification is made in the $j^{th}$ category, and so on. From this confusion matrix, sensitivity and specificity are then defined as
\begin{equation}
    \mathrm{Sen}_j = \frac{a_{j,11}}{(a_{j,11}+a_{j,21})}, \; \mathrm{Spc}_j = \frac{a_{j,22}}{(a_{j,12}+a_{j,22})}.
\end{equation} 
Subsequently, the F1-score is computed as
\begin{equation}
    \mathrm{F1}_{j} = \frac{2 (\mathrm{Sen}_j)(\mathrm{Spc}_j)}{\mathrm{Sen}_j + \mathrm{Spc}_j}.
\end{equation}
Note that the F1-score is computed for each outcome category separately. This enables us to identify biases in forecasts with regard to the classes, if any. A high F1-score implies that the model can consistently forecast the correct outcome. 

One of the criticisms of the F1-score is that it gives larger weight to smaller classes and favours models with similar sensitivity and specificity. It is imperative to use another criterion to evaluate the predictive accuracy of the models. Following the discussions by \cite{czado2009predictive} and \cite{kolassa2016evaluating} who pointed out the need to take into account the probability with which the forecast is made in similar problems, we are going to use the Brier score. It is a widely used scoring rule for multi-class prediction problems with mutually exclusive classes. The reader is referred to the works by \cite{brier1950verification} and \cite{murphy1973new} for the definition and related discussions on the Brier score. The scoring mechanism takes into account the probabilities of classification and compares them against the observed outcome. Recall the predicted probabilities in \cref{eq:predicted-prob} and let $\hat{p}_{ir}$ be the value corresponding to the $r^{th}$ category, for the $i^{th}$ match in $\mathcal{S}_{te}$. Then, the Brier score for the test set is given by
\begin{equation}
    \mathrm{Brier ~ score} = \frac{1}{m}\sum_{i=1}^{m} \sum_{r \in R} \left(\hat{p}_{ir} - \bm{1}(Y_i = r)\right)^2.
\end{equation}

Evidently, a lower Brier score implies a better predictive performance of the model. As mentioned before, these evaluation criteria will be reported for the competing models for every minute $t \in \Gamma$ of the match.

\section{Data}\label{sec:data}

\subsection{Description}

In this article, we use the data from English Premier League (EPL) matches from the 2008-09 season to the 2015-16 season. It is extracted from the European Soccer Database (ESD), which is available on Kaggle (link: \url{https://www.kaggle.com/datasets/hugomathien/soccer}). 

EPL is the top division in the English soccer system. Every season, 20 teams play in a double round-robin format, and in the end, the bottom three teams are relegated to the English Football League (EFL). To maintain the 20 team format, the top three teams from EFL are promoted to play in the EPL for the next season. Such a structure results in the total number of matches recorded over 8 seasons being $3040$. It is important to mention that it is not the same 20 teams playing in the EPL year on year, and consequently, the dataset has a total of 34 teams, each playing a varying number of seasons. One should note that the league matches finishing in tied scores by the end of regular time (90 minutes) do not go into extra-time or penalty shootouts. This enables an ordinal multinomial outcome with three categories. To analyze this dataset, we use a multinomial response variable to illustrate the match result for the team playing at home. As mentioned earlier, the covariates in the model are of two types: time-invariant covariates and time-varying ones. For the latter, remember that the dataset reports different types of events happening in every match, with their corresponding times of occurrence. We use eight such events in the model as time-varying covariates. These are goals, shots-on-goal, shots-off-goal, red cards, yellow cards, corners, crosses and fouls. Each event $k$ has an associated vector of covariates as defined in \cref{eq:pispec}. 

Regarding the time-invariant covariates, we consider the strength of the playing elevens for both teams. In order to define this, we rely on the ESD that records the overall rating for each player in the league, updated periodically based on their real-life performances on different parameters. These ratings consider 33 types of skills of the players, and the details can be found in the aforementioned link. Based on the information on the eleven players who start a game for each team, we compute the covariate depicting the overall strengths of the team. Specifically, the average of the ratings of the players in the starting eleven is computed based on the players' ratings at the beginning of the match. We point out that due to the absence of substitution data, the strength variable is assumed to be fixed over the course of the match, and thus, we classify it as a time-invariant covariate. Apropos to this point, note that the effects of specific opponents can also be estimated by introducing team-specific fixed effects, instead of the strength variables, in the model. We choose the latter to avoid the issues of over-fitting.

A brief discussion on the motivation behind the above choices is of the essence here. Previously, \cite{gonzalez2019effect} and \cite{gomez2018analysis} showed that the strength of a team based on its players and team rankings is useful in predicting the outcomes. Many other studies have used various events at an aggregate level as regressors for predicting match outcomes in soccer. \cite{liu2015match}, for example, used a generalized linear model to identify key winning indicators from 24 different event types from World Cup data. The authors identified shots, shots-on-target, tackles, red cards and crosses as important events which influence the result of a match. An interesting finding is the negative impact of crosses, which corroborates \cite{vecer2014crossing} who demonstrated through a multilevel Poisson regression model that crosses indeed have a negative impact on the goals scored by a team. Earlier, \cite{castellano2012use} used discriminant analysis to infer about various attack and defence attributes with regard to their effect on the result. Total shots and shots on target were found to have the greatest discriminatory power amongst the variables used. In more recent studies, \cite{ashimolowo2018econometric} investigated the association of crosses, corners, free kicks and the number of shots-on-goal with the outcome of a soccer match, whereas \cite{vcerveny2018effects} modelled the effect of a card on the goal-scoring rate of a team through a proportional hazard model.

\subsection{Exploratory analysis}\label{subsec:exploratory}

Before moving on to the main analysis, we find it imperative to present the descriptive statistics of the covariates used in the model. \Cref{tab:decriptive} displays the summary statistics of these variables for both home and away teams. We report the mean and standard deviation per game, along with the percentage of matches with no events of that type.

\begin{table}[!ht]
\centering
\caption{Summary of the covariates used in the analysis. Mean and standard deviation (SD) are computed per game, whereas the zeros column indicates the percentage of matches with zero cases.}
\label{tab:decriptive}
\begin{tabular}{ccccc}
     \hline
     & \multicolumn{2}{c}{\text{Home}} & \multicolumn{2}{c}{\text{Away}} \\
     \hline
    \text{Variable} & Mean (SD) & Zeros (\%) &   Mean (SD) & Zeros (\%) \\ 
    \hline
    Goal & 1.551 (1.312) & 22.829 & 1.160 (1.145) & 34.309 \\
    Shot-on & 6.684 (3.512) & 0.559 & 5.274 (2.936) & 2.072 \\
    Shot-off & 6.606 (3.091) & 0.428 & 5.212 (2.678) & 1.349 \\
    Red Card & 0.065 (0.254) & 93.717 & 0.097 (0.312) & 90.822 \\
    Yellow Card & 1.418 (1.170) & 24.243 & 1.802 (1.286) & 15.757 \\
    Corner & 10.196 (5.810) & 1.283 & 8.048 (5.080) & 2.368 \\
    Cross & 16.128 (7.609) & 0.066 & 12.448 (6.243) & 0.329 \\
    Foul & 10.691 (3.546) & 0 & 11.398 (3.687) & 0 \\
    Team Strength & 76.081 (3.784) & - & 75.843 (3.871) & - \\
    \hline
\end{tabular}
\end{table}    

We observe that the average number of goals scored by the home teams is slightly more than that by the away teams. This home advantage can be further illustrated by the fact that a home team is held scoreless in about $23\%$ of the matches as opposed to $34\%$ for the away team. This difference is a result of more shots-on-goal being taken by the home team. Identical observations can be made about shots-off-goal as well. When it comes to corners, the home team is awarded two more corners than the away team on average. Average crosses per the game trend in the same direction with the home team averaging four extra crosses. However, one may note that the strength variable, as expected, does not appear to be significantly different between the home and the away teams.

Coming to the disciplinary covariates, red cards are found to be infrequent events. In EPL, a red card is shown less than once in every ten matches on average. Comparing the raw per-game numbers, we observe that the away team is 1.5 times more likely to get a red card than the home team. Although the difference in yellow cards per game for the home and away teams is much smaller, the event of no yellow cards being shown in a match happens in nearly $10\%$ more matches for the home team as compared to the away side. In terms of the number of fouls, we observe a similar trend but to a lesser extent. One can possibly attribute these differences to refereeing bias towards the home team (\cite{boyko2007referee}).

\begin{figure}[!h]
    \centering
    \caption{Aggregate number of events (computed for 3040 matches in the dataset) of different types over the course of a match for home and away teams.}
    \label{fig:summary-covariates}
    \includegraphics[width = 0.8\textwidth,keepaspectratio]{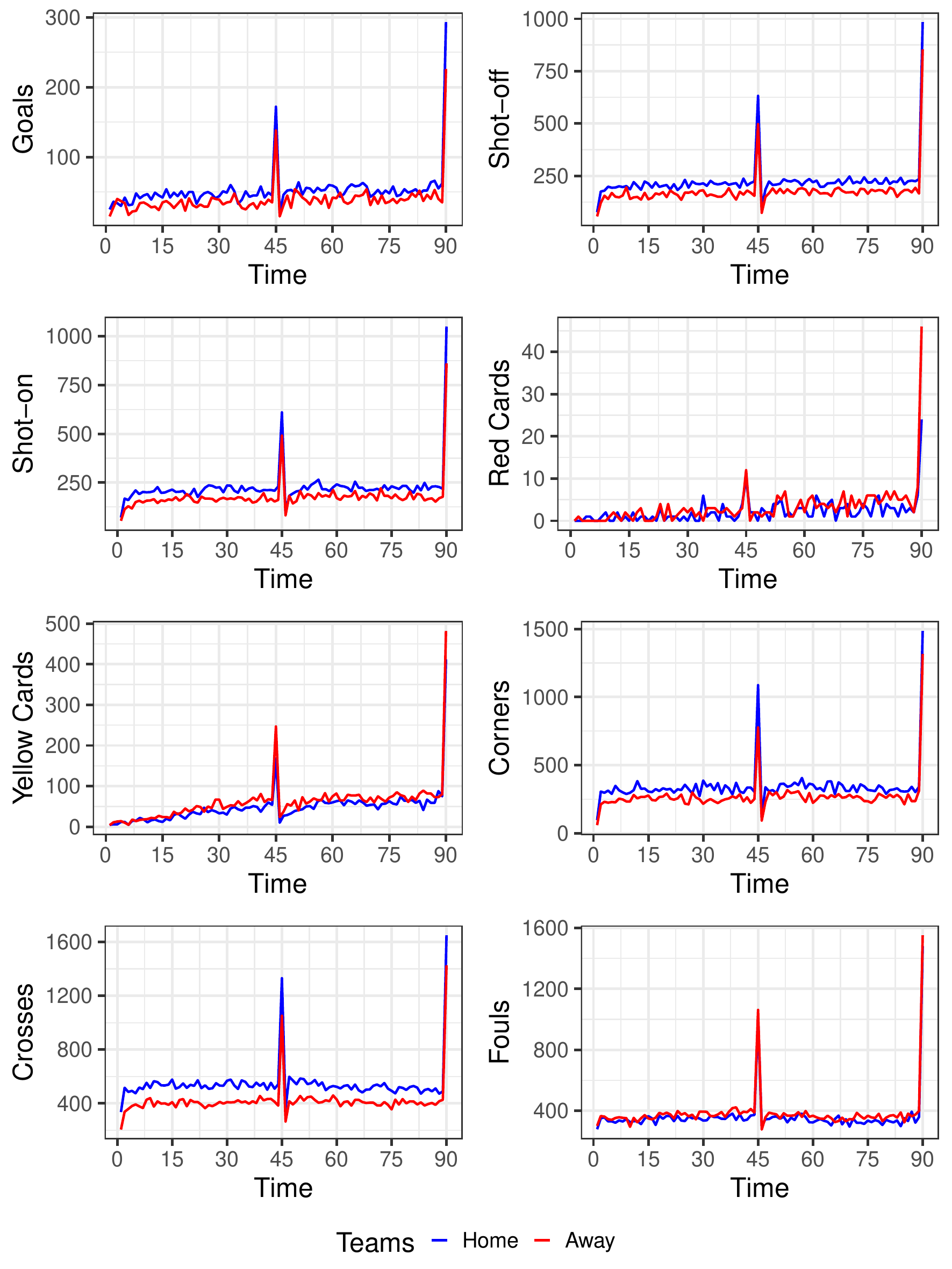}
\end{figure}

We further illustrate the nature of the time-varying covariates in the proposed model. To that end, \Cref{fig:summary-covariates} provides valuable insights into the aggregate number of events over the course of a match for both teams. In these plots, the total number of events for each type, computed from the entire dataset of 3040 matches, are presented. 

A striking feature for all the events is the spike observed at the $45^{th}$ minute and the $90^{th}$ minute. This phenomenon is related to the fact that stoppage time is provided to compensate for any delays that may have occurred during the preceding half, and during this time typically all teams tend to play with high intensity. We also want to point out that in this analysis, any event occurring in stoppage-time is reported corresponding to the last minute of the half, that is with the $45^{th}$ or the $90^{th}$ minute as appropriate, and therefore we are unable to assess any potential impact of the stoppage time on the outcome variable. 

Among the specific covariates, the pattern of the number of cards given as a function of time appears to be interesting. We observe that the frequency of yellow cards being handed out significantly increases as a match progresses. A similar increasing trend is observed in the case of red cards, albeit at a much slower rate. The difference in means between home and away teams for red cards can largely be attributed to the spike of the red cards given to the away team in stoppage time. However, no such trend has been found in cases of fouls, potentially hinting at the referees becoming stricter in their rulings, or a higher degree of aggressive play by both teams towards the end of the matches. We observe a systematic difference in the number of crosses and shots taken over time. This consequently results in an increased difference in the number of corners and the goals scored between the home and away teams. Interestingly, these differences are constant over time, but as we shall see in the main analysis, they are expected to have a differential impact on the final outcome of the game.

\section{Results and Discussion}\label{sec:results}

\subsection{Primary results}\label{subsec:primary-results}

For the main analysis, the focus is on understanding the predictive accuracy of the proposed methodology. In order to assess that, we split the dataset -- $90\%$ of the data are used for training the model, while the remaining $10\%$ are used as the test data. The results reported in this section correspond to this split, whereas in \Cref{subsec:robustness} we shall check the robustness of the algorithm by considering different training and test datasets. All calculations are carried out using RStudio Desktop (R version 4.3.1) using a 16-core 32GB processor. The codes for running the main algorithms are made publicly available on a GitHub repository maintained by the first author.  The library \texttt{future.apply} is used for parallel computing while \texttt{e1071} and \texttt{ranger} are used for the competing models. Note that the parallelization is used to reduce the computational time for the estimation of $90$ independent models, one for each time-point.

\begin{figure}[!h]
       \centering
       \caption{GW statistic (for the two key parameters) used to infer about convergence and decide the burn-in period for all 90 models.}
       \includegraphics[width = 0.8\textwidth,keepaspectratio]{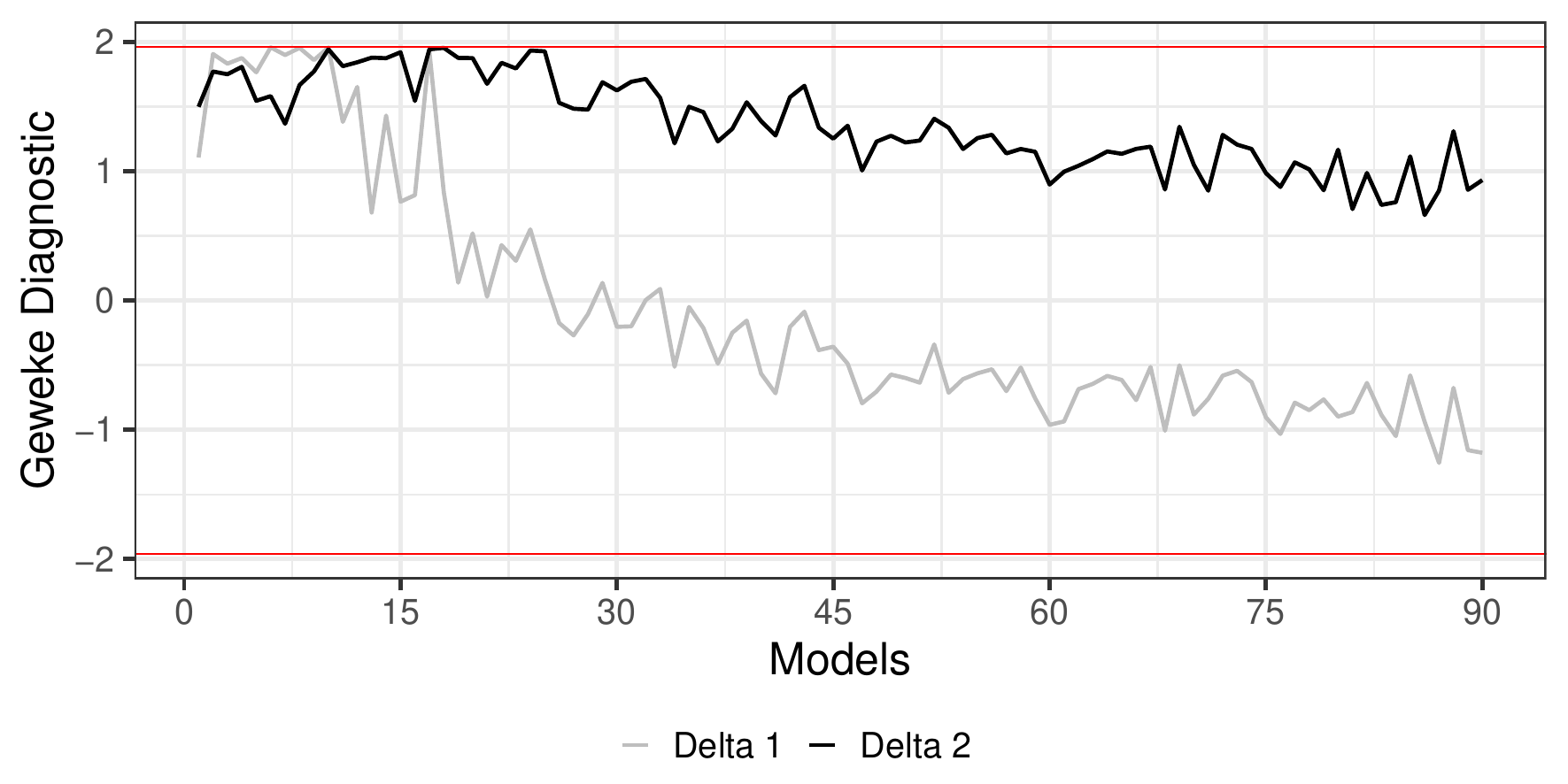}
       \label{fig:convergence}
\end{figure}

The prior specifications and the Gibbs sampling steps are detailed in \Cref{sec:methodology}. As stated there, the GW statistic is used as a diagnostic test for convergence. In order to obtain the posterior sample for the model parameters, we iterate until the GW statistic is below the threshold for all models which we classify as the burn-in period. By implementing thinning we then draw samples from the posterior after convergence. We first present the convergence results for all 90 models. A value between $-1.96$ and 1.96 for the GW statistic is an indicator of convergence. It is useful to note that we have specified a covariance structure which governs the posterior behaviour of $\bm{\nu}$. Hence, the convergence of the Gibbs sampler is evaluated based on the GW statistic for $\delta_1$ and $\delta_2$. \Cref{fig:convergence} shows the GW statistic for both $\delta_1$ and $\delta_2$ for every model post convergence, and we can ascertain that the convergence has been reached by all models. For all these models, the number of iterations required for achieving convergence ranges between 11,000 to 200,000, with a mean number of iterations being around 30,000. It is also observed that the convergence is faster for models with more information, i.e., for models at later time points. We acknowledge that the speed of convergence is impacted by the size of the data as well as suitable prior specifications, in line with the observations in similar Bayesian estimation problem for multinomial probit models by \cite{imai2005bayesian}. 


We restate that the goal is to forecast the match result in-game, that is when minute-by-minute data for the covariates are recorded in real-time. In line with this idea, we estimate the model with the time-varying covariates taken up to the $t^{th}$ time point for each $t \in \{1,2,\hdots,90\}$. We further forecast the match results for the test data for every such model using the mean of the posterior predictive distribution. Following the aforementioned notations, we use $\hat{\Pi}_{i}^{(t)}$ to obtain the predicted outcome of $i^{th}$ game when $t$ minutes of the match has passed, that is, the covariate information is available upto time $t$. 

\begin{figure}[!h]
       \centering
       \caption{Effect of each event on the outcome of the match captured at every time-point for both teams. The estimates displayed herein are from the model with complete match data and scaled by their respective standard errors.}
       \includegraphics[width = 0.8\textwidth,keepaspectratio]{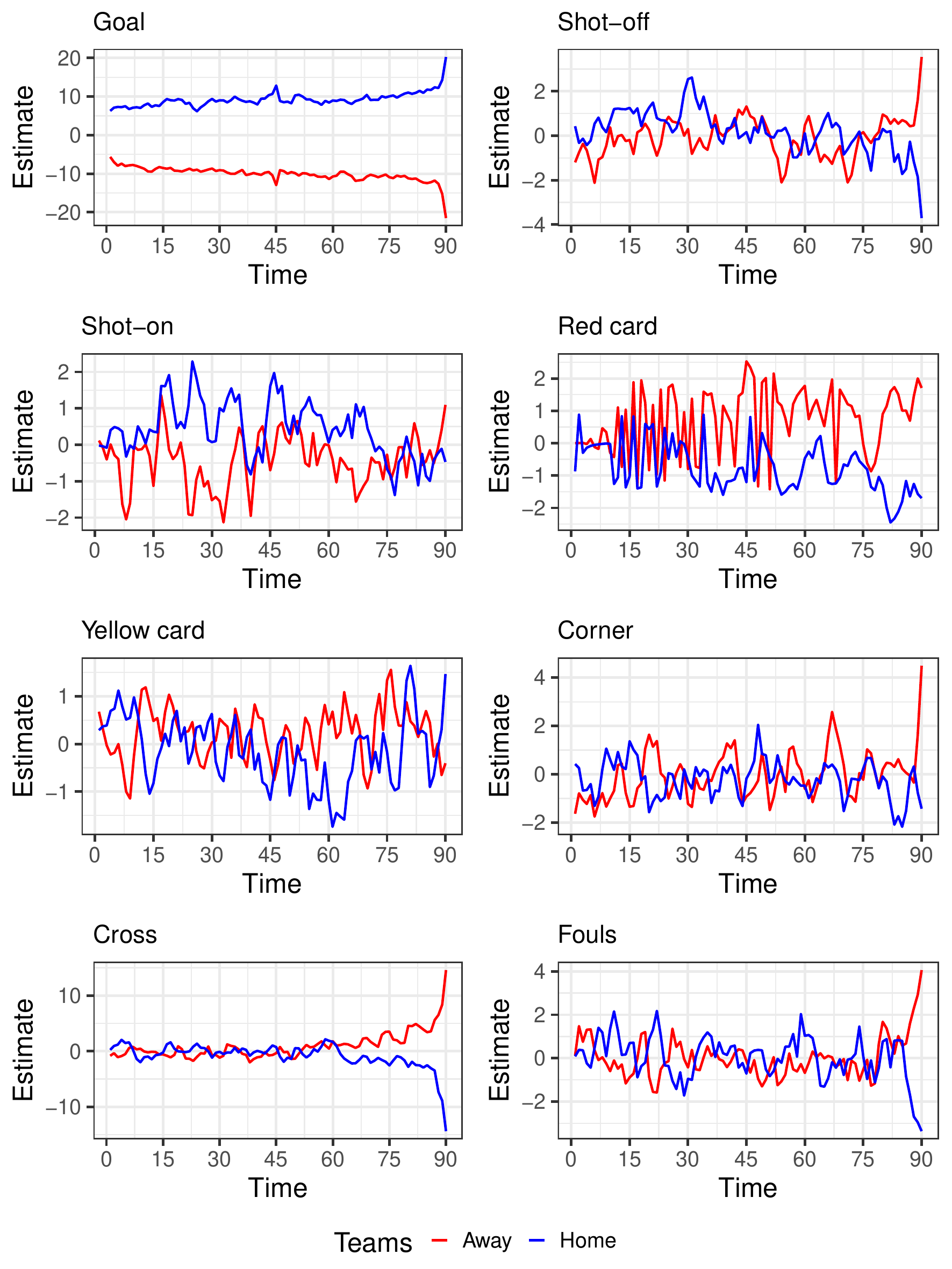}
       \label{fig:betas}
\end{figure}

We first examine the posterior estimates of the covariate effects in the model. From the posterior samples obtained from the proposed algorithm, we find the posterior means and the credible intervals for each parameter. In the case of the time-invariant covariates, the impact of the team strength is found to be substantial for both home and away teams. The coefficient of the home team is obtained to be 1.69, with the 95\% credible interval being $(1.17, 2.19)$. The same for the away team are $-1.32$ and $(-1.74, -0.83)$, respectively. As expected, one can see that the effect of home strength and away strength are relatively equal and of opposite signs. Next, in \Cref{fig:betas}, we demonstrate the temporal variation in the impact of various events on the probability of winning for the home team. A positive value implies an increase in the win probability by more occurrences of the event at that time-point, whereas a negative coefficient implies otherwise. In the figure, for ease of interpretation, all coefficients are scaled by the respective standard errors.

Goals are naturally the best indicator of the outcome of a match, as can be gleaned from the figure. We can observe an increasing effect of goals scored in the latter stages of the match as compared to the initial minutes. This phenomenon can be rationalized by the fact that a team has less time to respond when a goal is scored towards the end. The findings align with the results from \cite{castellano2012use}, \cite{vcerveny2018effects} and \cite{rocha2021influence}. Next, the variables shots-on-goal and shots-off-goal exhibit marginal influence on the outcome of a soccer match. Their impact is typically found to be negligible throughout the timeline. This can be attributed to the inclusion of goals scored as an event, which might capture most of the variability explained by shots-on-goal and shots-off-goal.

When it comes to cards, both red and yellow cards are usually negatively associated with the outcome. Specifically, a red card for any home is found to have a detrimental effect on their chances of winning, making it the second most crucial event in a soccer match. It is noteworthy that red cards, akin to goals, have a significant effect on the result over the course of the match; whereas yellow cards generally do not render deciding impact. With regard to temporal fluctuations, a red card awarded to the opponent in the latter stages of the match is more advantageous for the home team as compared to the away team. 

Corners, though deemed advantageous in a soccer match (\cite{ashimolowo2018econometric}), do not prove to be as effective in forecasting the outcome. Our findings are contrary to the common notion that corners are significant factors when awarded in the dying stages of a match. This is especially true in the $90^{th}$ minute and stoppage time where getting a corner does not impact the outcome in any significant manner. A contrasting behaviour is observed in the effect of a cross, which is conventionally a long-range pass made towards the opponent's goal from the two sides of the field. For the majority of the match, the influence of a cross on the outcome is discovered to be negligible, but after the $75^{th}$ minute, this changes. As time winds down, teams tend to be more aggressive and push for a goal. This usually leads to more long balls and crosses, and an inaccurate cross may result in a decisive counterattack on the other end, which has been hypothesized by other researchers as well \citep[][for example]{lepschy2020success}. Interestingly, this phenomenon is captured by the model which estimates a significant negative impact of crosses towards the end of the match for both teams. The negative influence of crosses keeps on increasing till the end. It is imperative to recall that this observation conforms with the results of \cite{vecer2014crossing} and \cite{liu2015match}, as discussed earlier.

\begin{figure}[h!]
    \centering
    \caption{\textcolor{black}{F1-score and Brier score (averaged over the test set) for different models, with respect to their within-game forecasting accuracy as a soccer match progresses}}
    \includegraphics[width = 0.9\textwidth,keepaspectratio]{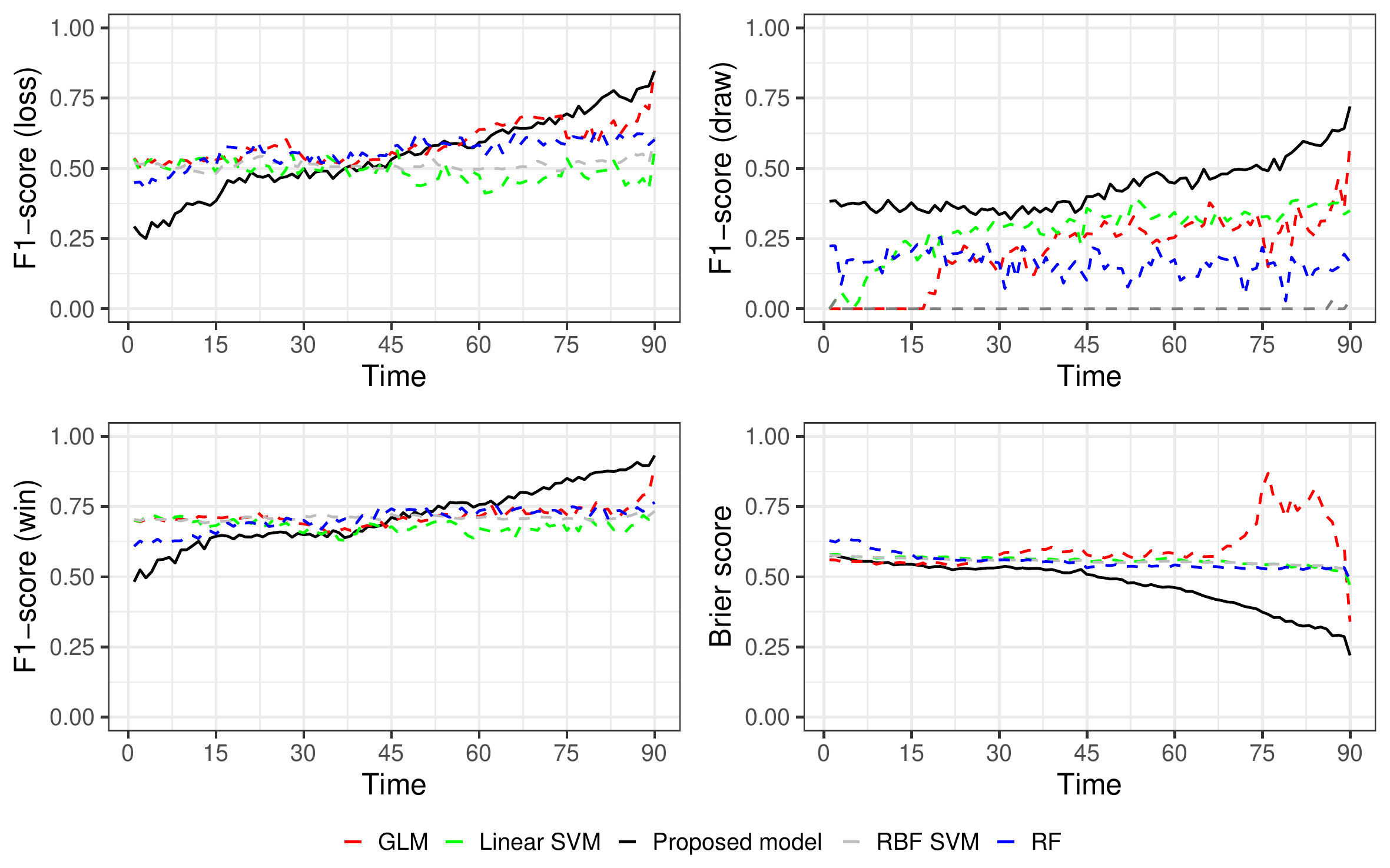}
    \label{fig:F1_Brier}
\end{figure}

We now move on to the results of the predictive accuracy, the main focus of this article. The F1-scores and the Brier score, as defined in \Cref{subsec:comparison}, are displayed in \Cref{fig:F1_Brier}. These accuracy metrics are computed as averages based on the predicted outcomes at each time point of the matches in the test set. We report the values for our method, along with the same for each of the four competing models.

F1-score, as discussed before, is determined for every outcome category independently. When the match result is a loss for the home team, we observe a distinctly higher F1-score for our model as the match unfolds. A similar trend is observed for the wins as well. Contrary to that, if the final outcome is a draw, the proposed method outshines its competitors throughout the time frame of the match. The linear SVM is the second-best-performing model in such cases.

To elaborate on the behaviour of the F1-scores thus obtained, one may argue that the competitor models weigh the effects of the goals in a much more severe manner as compared to the proposed model. Our model does not rely that heavily on goals and incorporates the effect of other events while forecasting. As a direct consequence of this, the proposed model performs much better than all the contender models while predicting drawn games. Also, in games where the scores are level, other covariates, such as team strength, home advantage, and player performance, are likely to have a greater impact on the forecast outcome which plays to our models' strengths. We emphasize that the aversion to relying heavily on the number of goals scored is unique to our model. This effect is further reflected upon in the case studies in \Cref{subsec:specific-examples}. Moreover, due to the availability of more covariate information, it is observed that in the latter stages of the match, our model provides uniformly better predictions than the rest. 

A look at the plot for Brier score tells a similar story. We emphasize that this scoring rule incorporates the predicted probability of the outcome for model evaluation. Generally, lower values are recorded by our method, but the scores are still comparable until the end of the first half, following which the Bayesian approach stands out significantly from its competitors until the end of the match. Akin to the earlier argument, we believe that the proposed model's performance is considerably better during the second half as it utilizes more available data in a better way. A peculiar observation can be made about the behaviour of GLM. Even though its F1-score does not display major irregularities, a significant spike in the Brier score is observed during the latter stages, likely due to its over-dependence on the number of goals. Overall, we can conclude that our model registers greater accuracy with more certainty than the contenders. 

In an attempt to understand the performances in a better way, we next report the sensitivity and specificity for each model captured at 15 minute intervals in \Cref{tab:sensitivity-specificity}. The primary metric of interest here is sensitivity and how it interplays among various match outcomes for every model. Linear SVM  outperforms all in terms of win sensitivity but has a significantly lower loss sensitivity. Furthermore, draws are not predicted by L-SVM since it performs binary classification. A behavioural pattern similar to L-SVM is observed for some of the other benchmark models. For instance, GLM lacks predictive power when it comes to draws; while RF demonstrates an inherent bias towards predicting an outcome of decisive nature. Consequently, the latter method exhibits the highest win sensitivity but the lowest draw sensitivity. Moreover, both R-SVM and GLM have decreasing win recall over time, making them less reliable at later stages of a match. Notably, all benchmark models underperform in terms of draw sensitivity, a result which gets reflected in \Cref{fig:F1_Brier} as well. It is important to note that for a model to be practically usable, achieving a balance in sensitivity across all outcome categories is crucial. Compared to the other models, it is evident that the proposed approach maintains this balance most effectively. A striking observation is that this balance increases over time, with both the metrics improving along the course of a match, which is not visible in any other method. 




\begin{table}[!ht]
    \centering
    \small
    \caption{Sensitivity and specificity captured at specific time points, corresponding to different outcome categories, for the proposed Bayesian model (PBM), generalized linear model (GLM), linear SVM (L-SVM), SVM with radial basis functions (R-SVM) and the random forest (RF).}
    \label{tab:sensitivity-specificity}
    \begin{tabular}{c|c|ccccc|ccccc}
    \hline
    Outcome & Minute & \multicolumn{5}{c|}{Sensitivity} & \multicolumn{5}{c}{Specificity} \\
    & & PBM & GLM & L-SVM & R-SVM & RF & PBM & GLM & L-SVM & R-SVM & RF \\
    \hline
    Win & 15th   & 0.603 & 0.849 & 0.897 & 0.732 & 0.720 & 0.766 & 0.500 & 0.354 & 0.614 & 0.551\\
        & 30th   & 0.589 & 0.746 & 0.925 & 0.651 & 0.801 & 0.778 & 0.607 & 0.392 & 0.690 & 0.551\\
        & 45th   & 0.637 & 0.712 & 0.931 & 0.646 & 0.877 & 0.823 & 0.696 & 0.399 & 0.766 & 0.550\\
        & 60th   & 0.678 & 0.719 & 0.904 & 0.630 & 0.863 & 0.873 & 0.728 & 0.405 & 0.772 & 0.551\\
        & 75th   & 0.781 & 0.733 & 0.897 & 0.603 & 0.883 & 0.918 & 0.722 & 0.411 & 0.804 & 0.538\\
    \hline
    Draw & 15th   & 0.555 & - & - & 0.206 & 0.159 & 0.545 & - & - & 0.830 & 0.896 \\
        & 30th   & 0.476 & 0.095 & - & 0.317 & 0.111 & 0.593 & 0.883 & - & 0.772 & 0.934\\
        & 45th   & 0.555 & 0.270 & - & 0.429 & 0.063 & 0.643 & 0.804 & - & 0.747 & 0.950\\
        & 60th   & 0.651 & 0.254 & - & 0.428 & 0.111 & 0.680 & 0.805 & - & 0.722 & 0.959\\
        & 75th   & 0.635 & 0.270 & - & 0.444 & 0.143 & 0.743 & 0.805 & - & 0.668 & 0.959\\
    \hline
    Loss & 15th   & 0.253 & 0.547 & 0.421 & 0.453 & 0.505 & 0.952 & 0.765 & 0.852 & 0.813 & 0.785\\
        & 30th   & 0.379 & 0.547 & 0.474 & 0.473 & 0.558 & 0.909 & 0.775 & 0.871 & 0.809 & 0.809\\
        & 45th   & 0.442 & 0.537 & 0.453 & 0.495 & 0.558 & 0.904 &  0.823 & 0.856 & 0.818 & 0.828\\
        & 60th   & 0.516 & 0.631 & 0.453 & 0.442 & 0.568 & 0.914 & 0.842 & 0.833 & 0.809 & 0.828\\
        & 75th   & 0.632 & 0.611 & 0.463 & 0.484 & 0.547 & 0.928 & 0.842 & 0.828 & 0.852 & 0.852\\
    \hline
    \end{tabular}
\end{table}

To gain further insights into the model forecasts, we capture the Brier scores after every 15 minute intervals with respect to the difference in goals (home minus away) observed for the matches at the end of these intervals. We observe that the lowest Brier scores aka best accuracy were obtained for matches with a goal difference bigger than one. Naturally, an instance of zero goal difference is the most difficult to predict for. It is evident from the decreasing Brier scores that the model performance improves over the course of a match. An interesting observation is the slight increase in Brier scores from the $15^{th}$ minute to the $75^{th}$ for unit goal difference categories. This is likely to be the cumulative effect of time-varying covariates other than goals. This might also be an indication that goals scored in the early stages of a match provide better predictability than accounted for. It is possible that a different covariance structure than the one used in the model might be more effective in capturing this effect.


\begin{table}[!h]
    \centering
    \small
    \caption{Brier scores (with respect to goal difference) corresponding to the final outcomes based on predictions at specific time points.}
    \label{tab:goal_diff_brier}
    \begin{tabular}{c|ccccccc}
    \hline
     Minute & \multicolumn{7}{c}{Goal difference at the time of prediction} \\
     & $\leqslant -3$ & $-2$ & $-1$ & 0 & 1 & 2 & $\geqslant 3$ \\
     \hline
    15th & - & 0.034 & 0.237 & 1.36 & 0.247 & 0.015 & - \\
    30th & - & 0.056 & 0.414 & 0.921 & 0.440 & 0.058 & 0.011 \\  
    45th & 0.004 & 0.135 & 0.411 & 0.768 & 0.439 & 0.111 & 0.033 \\  
    60th & 0.051 & 0.100 & 0.395 & 0.754 & 0.388 & 0.138 & 0.075 \\    
    75th & 0.055 & 0.130 & 0.496 & 0.627 & 0.341 & 0.148 & 0.102 \\    
    \hline
    \end{tabular}
\end{table}


\subsection{Robustness of the analysis}\label{subsec:robustness}

To further support our previous results, we conduct a few robustness checks. One way to evaluate the heterogeneity in model forecasts is to compare the forecast accuracy based on the strength of a team. In that aspect, we split the set of EPL teams into two groups -- `Big 6' (consisting of Arsenal, Chelsea, Liverpool, Manchester City, Manchester United, Tottenham) and the rest. These `Big 6' teams are known for their large payrolls and fan-bases across the world. Our objective is to identify if the predictive accuracy is different for these teams, as compared to the conventionally weaker teams. \Cref{fig:big6} below shows the Brier scores for the matches corresponding to these two groups, split further according to who they played against.

\begin{figure}[!h]
    \centering
    \caption{Brier scores as a function of time when the teams are classified into two categories, namely `Big 6' and others}
    \includegraphics[width = 0.9\textwidth,keepaspectratio]{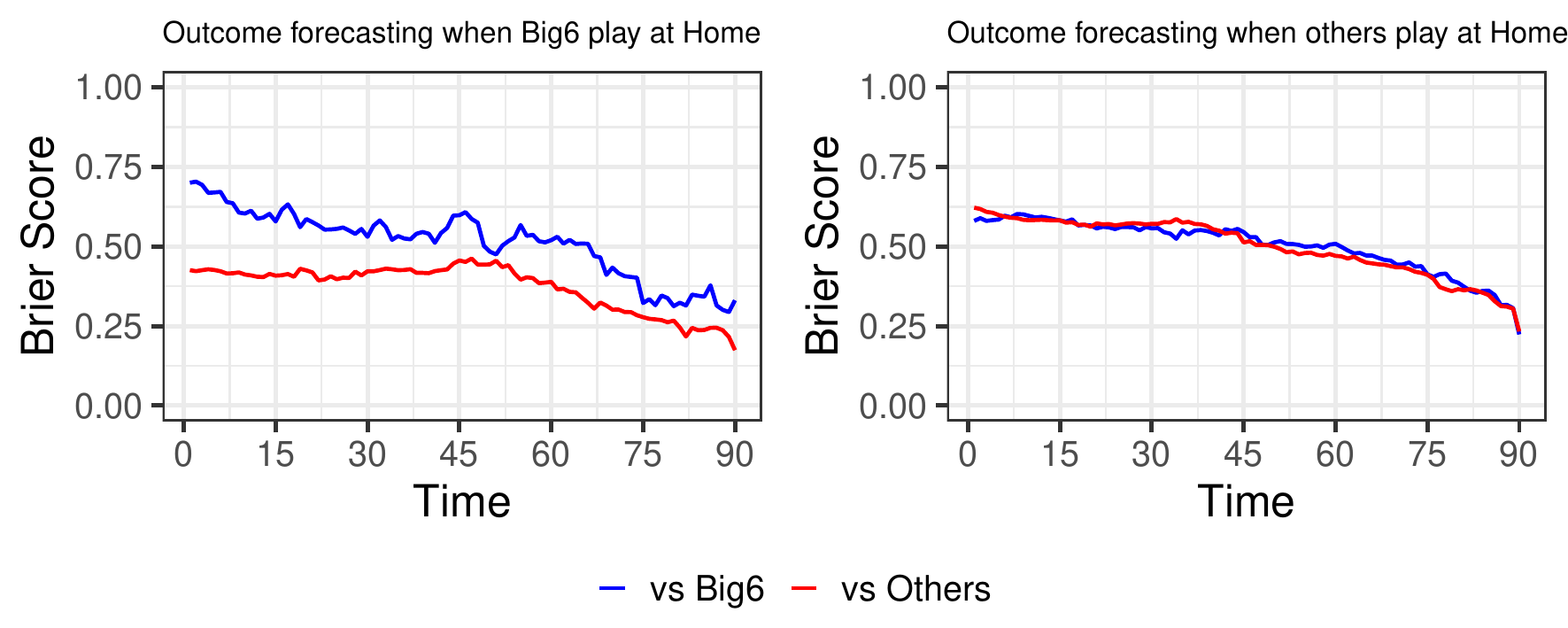}
    \label{fig:big6}
\end{figure}


Home advantage is a commonly observed phenomenon in soccer (\cite{staufenbiel2015home}). The `Big 6' clubs are the largest and most successful clubs in English soccer. One can hypothesize that they have a stronger home advantage than smaller teams, providing better predictability of the outcomes when these clubs play at home against the others. This is reflected in the lower Brier scores in such situations, especially in case of correctly predicting the wins of the `Big 6' (in fact, we observed that the brier scores of the predictions made at half-time corresponding to win is nearly four times better than the same corresponding to draw or loss of these teams). Interestingly, when two of the `Big 6' clubs play against each other, the average Brier score is found to be on the higher end during the first half, suggesting that the matches are unpredictable in these circumstances. As discussed in \Cref{subsec:primary-results} before, team strength plays a crucial role in forecasting match outcomes in the early stages. Due to the strength being comparable for both teams, the above phenomenon can be observed. For other teams playing at home, our model is robust in its accuracy irrespective of the opponents. Additionally, games involving the `Big 6' clubs against other teams at their home stadiums also show significantly better forecasting accuracy in general. This suggests that home advantage is an effective performance indicator for the bigger teams, but not as much for smaller market teams. After the 60th minute mark in these matches, we note that the metric is similar for all cases, thereby establishing the proposed model's robustness in this aspect.



\begin{figure}[!ht]
    \centering
    \caption{Brier scores computed after excluding a team's games from the training data. The test data comprises entirely of matches played by the excluded team.}
    \includegraphics[width = \textwidth,keepaspectratio]{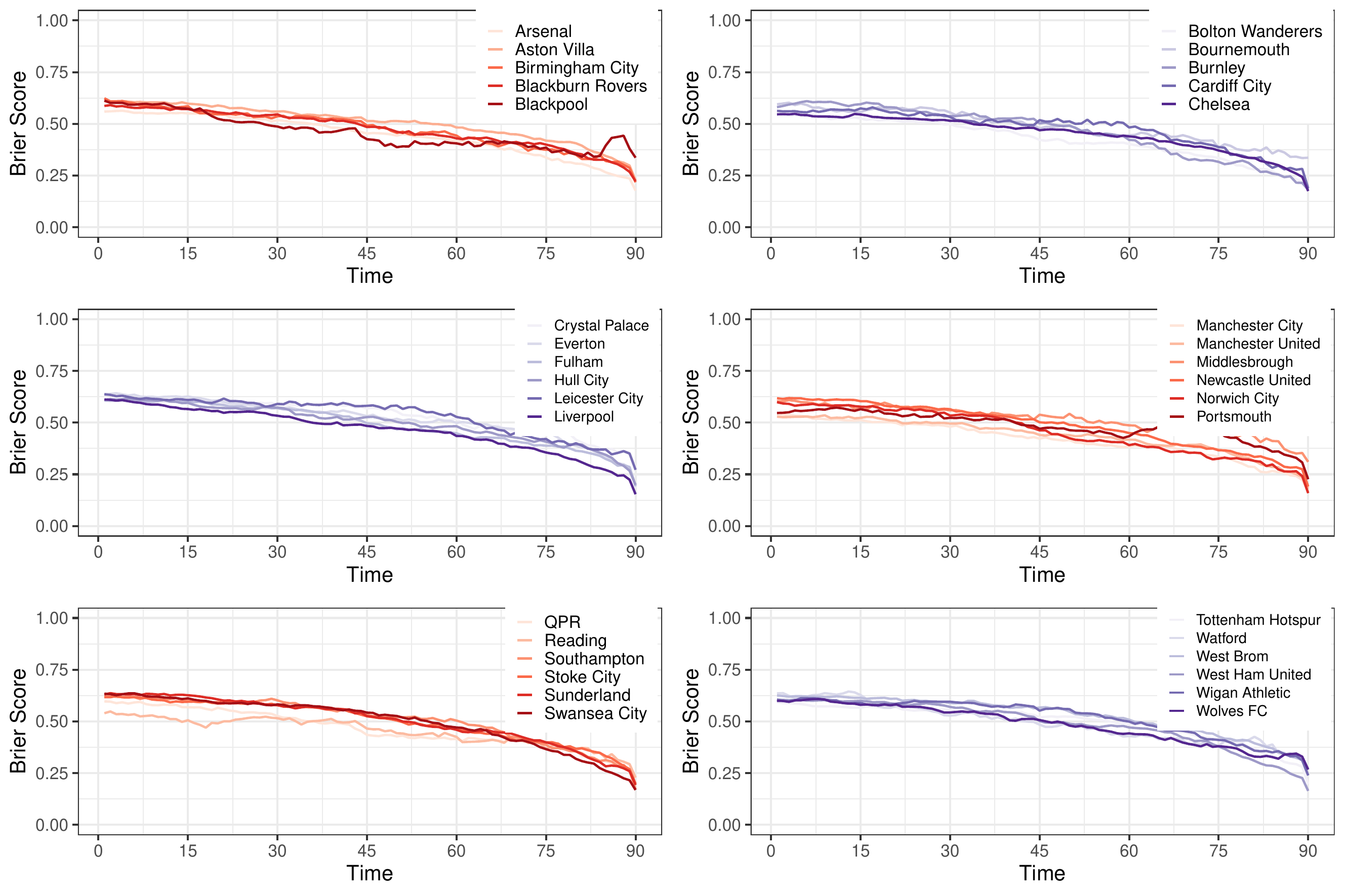}
    \label{fig:specific}
\end{figure}

Next, we look at a different perspective where we predict the outcomes of the matches for a certain team, by training the model on the dataset excluding all matches of that team. This exercise is performed for each of the 34 teams in the dataset, and we compute the average Brier scores, which are reported in \Cref{fig:specific}. 

It must be noted that the teams have different extents of variations in their year-end positions. For example, Manchester United is typically at the top table in all seasons, whereas Aston Villa is a team with an average year-end position of 13 with the highest and lowest positions being 7 and 17. Some teams fluctuate in and out of the league on a yearly basis due to relegation as well. The idea is to identify how well the model performs when forecasting for all these teams, even without using its data in the training set. We believe that a robust model should be able to forecast the results of across all scenarios. The plots in \Cref{fig:specific} indicate that the model registers comparable accuracy in predicting the outcomes of games for all teams. This suggests that the model effectively captures the key features of the teams and their performances, even in the presence of notable fluctuations from one year to the next. This is a significant result, as it implies that the model may be able to make reliable predictions for a wide range of teams and situations. A natural extension from a modelling perspective is the use of team-level fixed effects instead of the home-away classification we have currently used in the model. However, we observed that this gives rise to overfitting and consequently inferior results. Therefore, one will not be able to capture the effect of home advantage in such a model.

\subsection{Two specific examples}\label{subsec:specific-examples}

As a last piece of discussion in this section, we look at the performance of the proposed approach for two specific games of different flavours. For each game, the model is trained on the data of all other matches before it, and the in-game forecasting is done on a minute-by-minute basis. 

\begin{figure}[!ht]
    \centering
    \caption{Minute-by-minute forecast of the win probability for both teams during the match between Chelsea and Portsmouth.}
    \includegraphics[width = 0.8\textwidth,keepaspectratio]{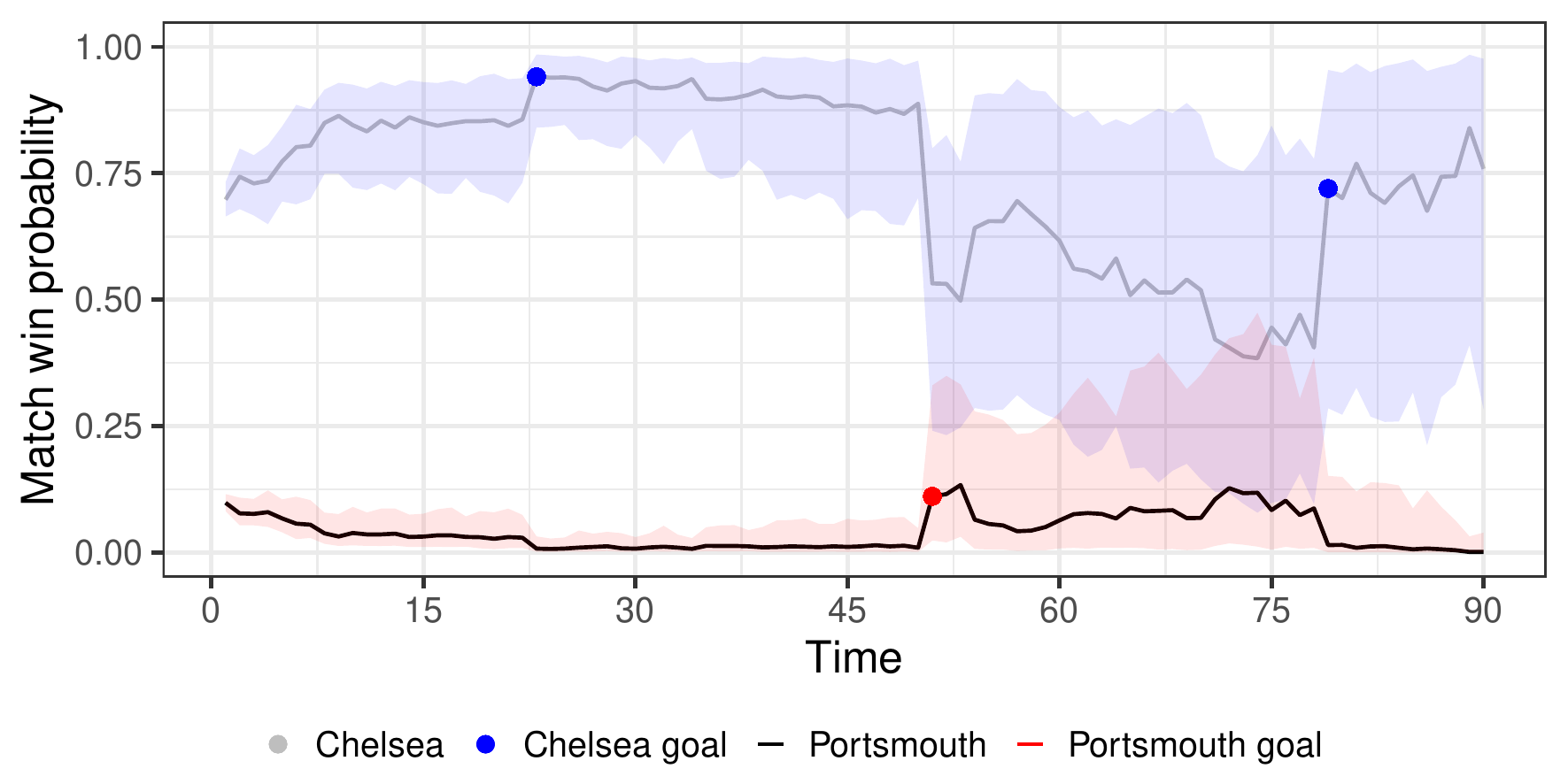}
    \label{fig:case1}
\end{figure}

The first game we discuss was played between Chelsea and Portsmouth, at the home ground of the former. It was a back and forth match with a goal towards the end, giving Chelsea the win. This case study demonstrates the ability of our model to accurately forecast the outcome by considering a variety of factors, including team ratings and time-varying covariates to understand the flow of the game. The within-game forecasting of the win probabilities and the 95\% credible intervals for the two teams is displayed in \Cref{fig:case1}. 

One can note that the final scoreline read 2-1 with the winning goal in the match being scored in the $79^{th}$ minute. Chelsea was the heavy favourite at the start of the match, both due to being a better team (with respect to team rating) and for playing at home. The predicted win probability at the start was around 75\%, and it improved for a few minutes in the beginning. A goal in the $21^{st}$ minute consolidates our forecast. Later, Portsmouth equalled the scoreline just after the start of the second half. It is indeed worth mentioning that the goals scored by the two teams are found to have differential impacts on the probabilities, potentially due to the other covariates used in the model. For instance, regardless of the even scoreline after the $50^{th}$ minute, our model still considers Chelsea to be the odds on favourites to win. Chelsea is predicted to emerge victorious with 50\% chance for the remaining of the match, and the result indeed goes in their favour at the end. Clearly, this case study serves as a captivating example of the importance of considering team strength in predictive modelling and the ability of our model to incorporate several factors accurately. We find it imperative to point out that the credible intervals for both teams are narrow during the early stages of the match. These get wider as the match progresses, especially after Portsmouth equalizes. This correctly reflects the uncertainty in the outcome as the scoreline is 1-1. A mix of crosses, corners and shots on goal by Chelsea resulted in large credible intervals persisting until the end of the match.

\begin{figure}[!ht]
    \centering
    \caption{\textcolor{black}{Contributory effect of all time-varying events on the probability of winning for Chelsea v/s Portsmouth. The first figure contains exclusively Chelsea events. The second figure contains exclusively Portsmouth events.}}
    \includegraphics[width = 0.8\textwidth,keepaspectratio]{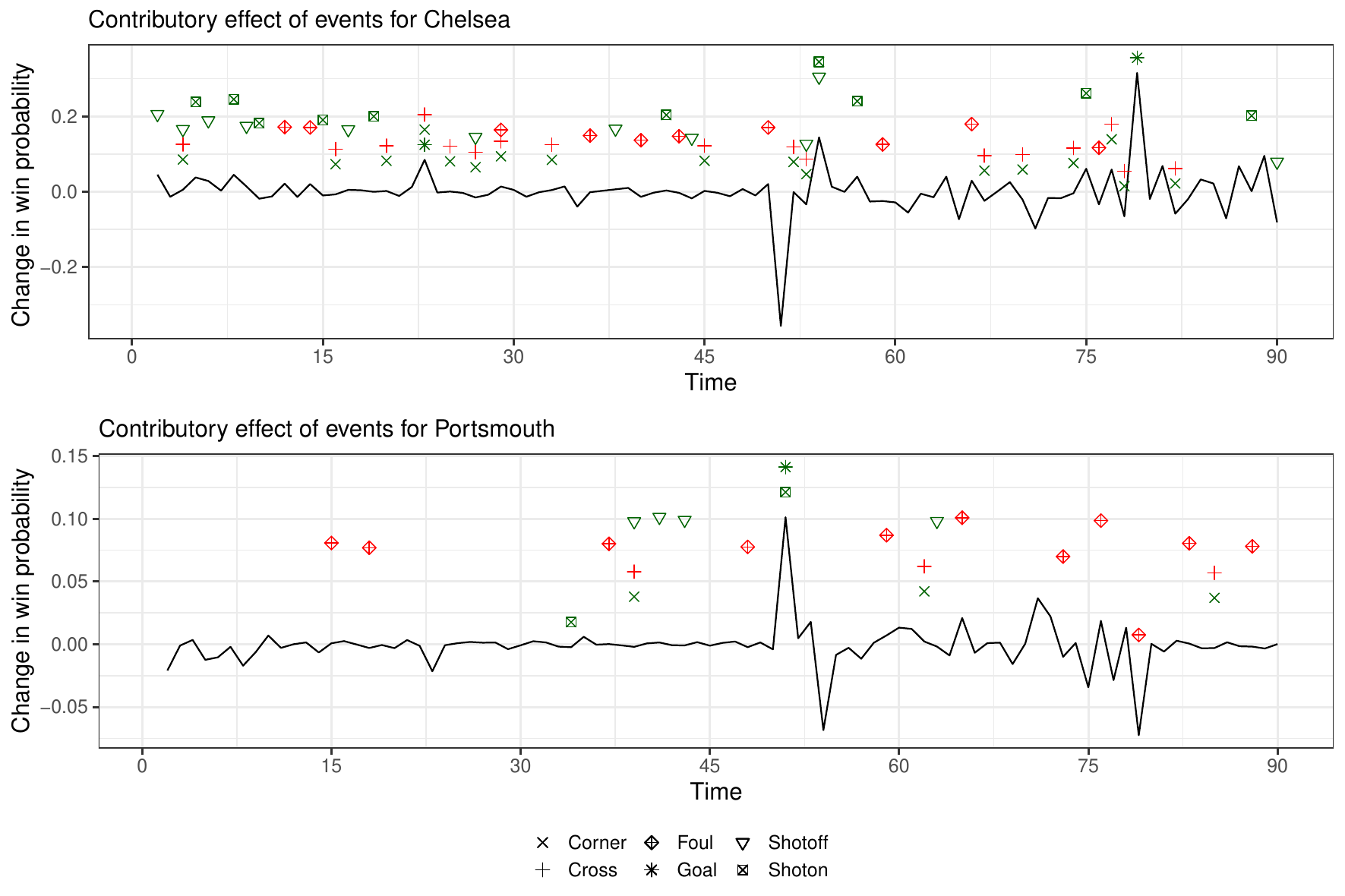}
    \label{fig:case1_contri}
\end{figure}

In \Cref{fig:case1_contri}, we further attempt to demonstrate the probabilistic effect of the time-varying events for the match. It is to be noted that the events plotted are only for those respective teams. It is quite interesting that events other than goals have a significant impact on the opponent's win probability but a comparatively muted effect on their own. For instance, a goal by Portsmouth contributes to the sudden drop in the win probability for Chelsea, but only a marginal increase in the same for Portsmouth. On the other hand, a host of shots-on and shots-off-goals in the initial minutes enhances the win probability for Chelsea. This can have profound managerial implications. By observing the effects these time-varying covariates have on the respective team's win probabilities, the managers can make in-game adjustments to counter the same.

\begin{figure}[!ht]
    \centering
        \caption{Minute-by-minute forecast of the win probability for both teams during the match between Bolton Wanderers vs West Bromwich Albion}
    \includegraphics[scale = 0.4]{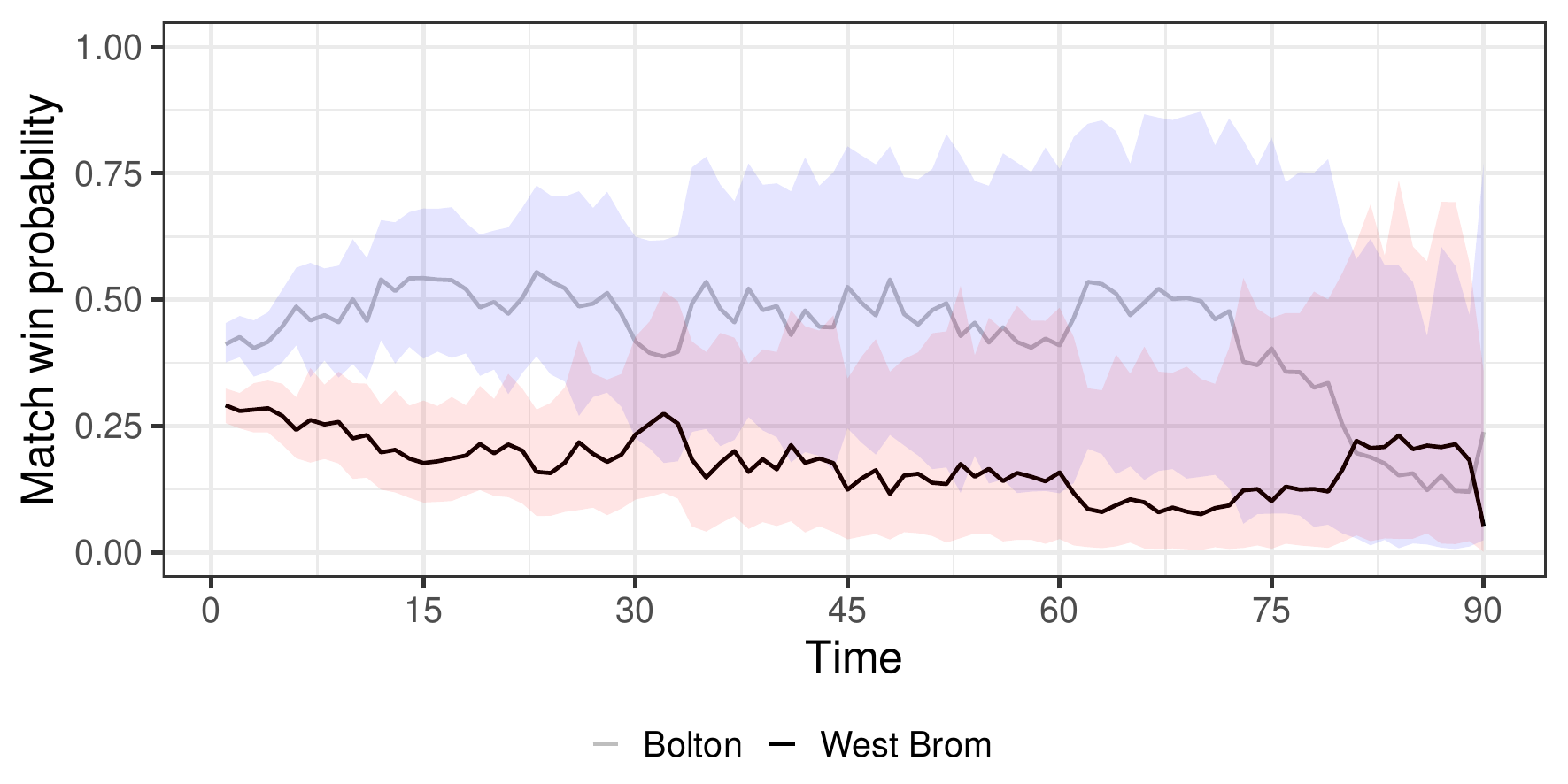}
    \label{fig:case3}
\end{figure}

\begin{figure}[!h]
    \centering
    \caption{Contributory effect of all time-varying events on the probability of winning for Bolton Wanderers vs West Bromwich Albion. The first figure contains exclusively Bolton events. The second figure contains exclusively West Brom events.}
    \includegraphics[width = 0.8\textwidth,keepaspectratio]{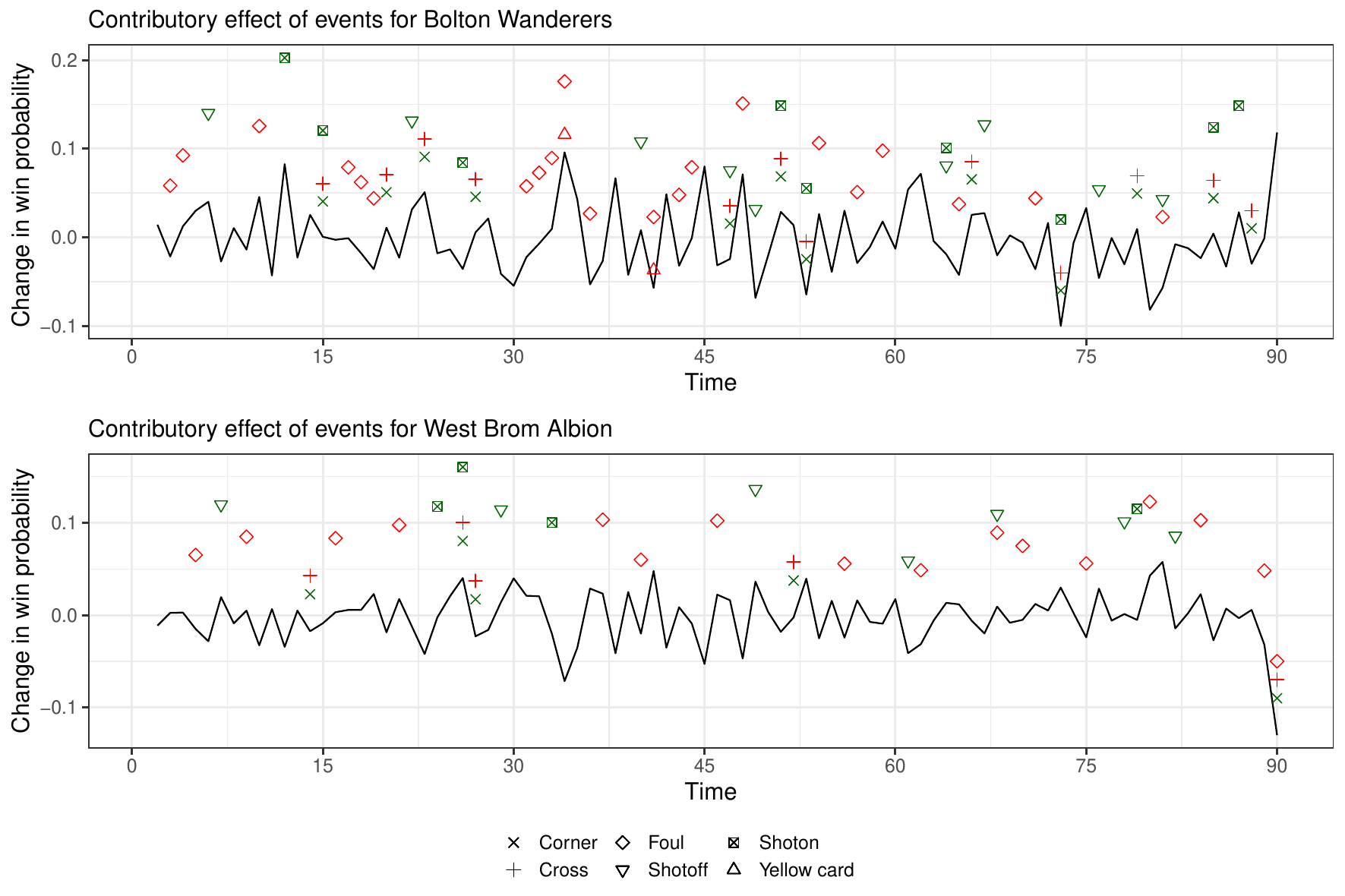}
    \label{fig:case2_contri}
\end{figure}

For completeness, we now dive deep into another case where our proposed model did not yield great results. The game was played between Bolton Wanderers and West Bromwich Albion, both teams typically being at the bottom of the league table. The score at the end of the $90^{th}$ minute is 0-0, a draw. As we can see in Figures \ref{fig:case3} and \ref{fig:case2_contri}, at the start of the game, our model gives a slight advantage to Bolton since it is playing at home. Then, a few shots-on-goal by Bolton pushed their win probability past $50\%$. This pattern persists for most parts of the match, with Bolton being considered the favourites until the $70^{th}$ minute. It is only after this that the most likely predicted outcome is a draw. As we illustrate in \Cref{fig:big6}, home advantage is not an effective performance indicator for smaller market teams, which contributes to the inaccurate forecast for the majority of the game. This case further reflects the results obtained in \Cref{tab:goal_diff_brier} regarding the difficulty in forecasting games with an equal scoreline.

We also notice that the credible intervals get wider as the match progresses, thereby reflecting the uncertainty in the forecast. We believe that this is primarily due to the teams being equally matched. Further, the absence of goals gives us a better insight into the effect of other time-varying covariates. We can observe that shots-on-goal and fouls have a substantial effect on the probabilities in the absence of goals. A peculiar observation is the sudden rise in win probability for Bolton in the dying moments of the match. As we can identify in the bottom panel of \Cref{fig:case2_contri}, a sequence of fouls and a cross by West Bromwich players towards the end may have provided this edge to the home team.

\section{Conclusion}\label{sec:conclusion}

\begin{figure}[ht!]
    \centering
    \caption{Example of minute-by-minute forecast for a particular match between Manchester United and Sheffield United, sourced from \url{dimers.com} (see \url{https://shorturl.at/fyILO}).}
    \includegraphics[scale = 0.8]{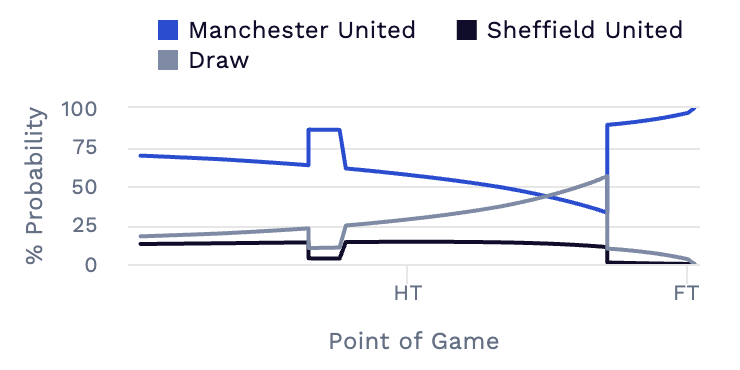}
    \label{fig:Dimers-in-game}
\end{figure}

In this article, we developed a Bayesian latent variable model for analyzing and forecasting soccer match outcomes in real-time. The modelling approach, to the best of our knowledge, is the first in academic literature to furnish real-time predictions in soccer via a Bayesian technique. As mentioned earlier, although several websites offer in-game probabilities for various matches, the lack of clarity and documentation behind their methodologies affect their practical utility from a managerial standpoint. Another advantage of the proposed model over these websites is its ability to quantify the uncertainty in the probabilistic forecast at each time point. As an illustration, we present \Cref{fig:Dimers-in-game} (sourced from \url{dimers.com}), which depicts the in-game probabilistic forecast during a recent EPL match between Manchester United and Sheffield, with the former winning by a scoreline of 2-1. A notable contrast between this and Figures \ref{fig:case1} or \ref{fig:case3} is the absence of credible intervals for the predicted probabilities. Moreover, noting the three jumps in the predicted probabilities in this case, it is apparent that the outputs are mainly affected by the goals scored; while our method is able to appropriately capture the effects of all types of significant events. Thus, we strongly believe that the proposed method, by virtue of its transparency and flexibility, can assist managers in formulating statistically sound strategies during a match. These aspects are crucial for any sort of decision making, be it from a spectator's perspective or that of a manager's.

In our application, we used the data from EPL matches across eight seasons, with minute-by-minute data of various events in the matches. The main variable of interest for each game was the outcome, modelled as an ordered multivariate random variable with three categories of response. We used a latent variable and two cut-offs on it as a proxy for the final outcome and modelled this latent variable using a linear functional form. The functional form involved time-variant and time-invariant covariates as well as their corresponding coefficient, along with a random error term. In our computations, we use a Gaussian prior for the random errors, the coefficient as well as the cut-offs, and a Dirichlet prior for the joint probabilities based on the cut-offs, to compute the step-wise conditional distributions as well as the posterior predictive distribution. It is critical to note that the Bayesian computations in our method are considerably fast due to the specific choices of the priors. However, an interesting future direction will be to relax the Gaussian assumptions and see how prior design may affect convergence in similar complex multinomial problems.

Coming to the performance in terms of forecast accuracy, we notice that our model provides insights into the effect of various events, such as corners, crosses and cards in soccer, on the outcome. Through various evaluation criteria, we find the model to be highly effective across different scenarios. We have seen that the results are robust for conventional big teams, as well as for teams with inconsistent performances. Furthermore, taking two specific examples, we demonstrate the effectiveness of the model in predicting the outcome well beforehand, a decisive goal is actually scored. Keeping that in mind, we strongly believe that the proposed methodology can be an extremely useful tool to maintain audience engagement in broadcasting soccer matches, or for the betting markets in real-time. A closely related area of research would be to understand the improvement in the win probability by potential substitutions. It can be achieved by including the player-level information in the model. Based on the within-match statistics and their deviation from the historical performances of the players, one may continuously compute the win-probability for different choices of replacements, which would in turn facilitate the coaches to devise a statistically sound substitution strategy during a match.

Continuing along the line, we would like to end the paper with a short account of other possible future extensions of this study. We note that the data did not include detailed information on the passing sequence, ball possessions, or substitutions. These aspects are expected to have a strong impact on the outcome of the game, but unfortunately, due to the lack of data, we could not add them in the current analysis. Another aspect we lose out on is team cohesion based on the interaction between individual players both in the same and opponent teams, which is also impacted by substitutions. Naturally, future implementation of the model can include these events in the framework and assess their impact on the match outcomes. Another possible improvement in the prediction accuracy can be to incorporate dependencies among different pairs of teams with respect to their previous outcomes. Similarly, one may consider the effects of the events from the home and away teams to be dependent, by relaxing the assumption in \cref{eq:beta-home-away-indep}. Utilizing appropriate structural forms in this context, $\Sigma_k^{(t)}$ in \cref{eq:beta-distribution} can be readjusted and then, it would be possible to develop a Bayesian algorithm in an identical fashion. Albeit it is expected to increase the computational burden for the Markov chains to converge, predictive ability may improve and such an idea is worth exploring in future works. Furthermore, in the current work, we do not consider self-exciting events and have considered different types of events to be independent of each other. It is not the ideal scenario, as given the occurrence of one kind of event, some other events may have higher chances of occurring. Addressing these different types of correlations in the model offers excellent scopes of future works to our method.

On a related note, another intriguing future direction is to assess the aspect of within-game momentum in soccer \citep[see][for some relevant discussions]{gauriot2018psychological,otting2021copula}. In order to do this, one can extend the proposed model to a time series setting, where the response is a multivariate observation indicating the occurrence of different types of events and the regressor set includes the information of the past events in the game. This framework will allow one to estimate the momentum effect of different types of events and can subsequently lead to improved forecasting capability of the final outcome as well. Last but not the least, we recall that the assumption of Gaussian priors on the coefficients associated with the real-time events might be construed to be restrictive. We plan to address this issue in future works, by considering a more general structure, possibly along with a semi-parametric approach to model the outcome of the game. 

One must recognize that the proposed methodology can be suitably modified and adapted to various interesting research problems as well. In other sports, such as basketball or cricket, a direct extension is natural and can offer an effective tool of within-match forecasting. Not only does this have the potential to be utilized in betting markets, but it can also be used to figure out an appropriate substitution strategy in basketball where the rules allow rolling substitution. In fact, for other problems involving ordered categorical response variables, one can develop a similar technique. For instance, in sports broadcasting industry, such an algorithm can help us determine the effects of various events in retaining a customer. One may also develop a similar Bayesian method to understand the impact of different types of events and announcements on how a firm performs in an economy.

\section*{Data availability statement}

The data used in this study are obtained from the publicly available European Soccer Database in Kaggle (link: \url{https://www.kaggle.com/datasets/hugomathien/soccer}). All codes for data extraction and for running the main algorithms are made publicly available on a GitHub repository maintained by the first author (link : \url{https://bit.ly/3Oo0SN8}).


\end{document}